\documentclass[aps,reprint,superscriptaddress,twocolumn,floatfix,showpacs,longbibliography]{revtex4-1}

\usepackage{graphicx,color}
\usepackage[utf8]{inputenc}
\usepackage{amsmath,amssymb,bm,amsfonts,dsfont,mathrsfs,amsthm}
\usepackage{braket}
\usepackage{txfonts,comment}
\usepackage[colorlinks=true,linkcolor=blue,citecolor=blue,urlcolor=blue]{hyperref}
\usepackage{soul}
\usepackage{enumitem}
\usepackage{multirow}
\usepackage{orcidlink}

\newcommand{\nn}{\nonumber \\}

\AtBeginDocument{%
    \newwrite\bibnotes
    \def\bibnotesext{Notes.bib}
    \immediate\openout\bibnotes=\jobname\bibnotesext
    \immediate\write\bibnotes{@CONTROL{REVTEX41Control}}
    \immediate\write\bibnotes{@CONTROL{%
    apsrev41Control,author="08",editor="1",pages="1",title="0",year="1"}}
     \if@filesw
     \immediate\write\@auxout{\string\citation{apsrev41Control}}%
    \fi
}%

\begin{document}

\title{Modular non-Hermitian topology and its application to critical sensing}\

\author{Saubhik Sarkar\,\orcidlink{0000-0002-2933-2792}}
\thanks{saubhik.sarkar@uestc.edu.cn}
\affiliation{Institute of Fundamental and Frontier Sciences, University of Electronic Science and Technology of China, Chengdu 611731, China}
\affiliation{Key Laboratory of Quantum Physics and Photonic Quantum Information, Ministry of Education, University of Electronic Science and Technology of China, Chengdu 611731, China}

\author{Chiranjib Mukhopadhyay\,\orcidlink{0000-0002-4486-9061}}
\thanks{chiranjib.mukhopadhyay@uestc.edu.cn}
\affiliation{Institute of Fundamental and Frontier Sciences, University of Electronic Science and Technology of China, Chengdu 611731, China}
\affiliation{Key Laboratory of Quantum Physics and Photonic Quantum Information, Ministry of Education, University of Electronic Science and Technology of China, Chengdu 611731, China}

\author{Abolfazl Bayat\,\orcidlink{0000-0003-3852-4558}}
\thanks{Corresponding author: abolfazl.bayat@uestc.edu.cn}
\affiliation{Institute of Fundamental and Frontier Sciences, University of Electronic Science and Technology of China, Chengdu 611731, China}
\affiliation{Key Laboratory of Quantum Physics and Photonic Quantum Information, Ministry of Education, University of Electronic Science and Technology of China, Chengdu 611731, China}
\affiliation{Shimmer Center, Tianfu Jiangxi Laboratory, Chengdu 641419, China}

\begin{abstract}

Non-Hermitian topological systems have attracted a lot of research activities in recent times, both theoretically and experimentally, due to their unique physical properties and association with open quantum systems.
We show that modular structures, where specific couplings at regular intervals take distinct values, enrich the unique topological attributes of these systems such as the non-Hermitian skin effect and the breakdown of conventional bulk-boundary correspondence.
These systems also possess the capability of displaying criticality-enhanced sensitivity for precision metrology.
We establish how the modular structure, in suitable parameter regime, enhances their sensing performance near spectral topological phase transitions and show that the enhancement can be achieved in multi-parameter estimation scenarios as well.

\end{abstract}

\maketitle


\section{Introduction} 

\begin{figure*}[t]
\centering
\includegraphics[width=\linewidth]{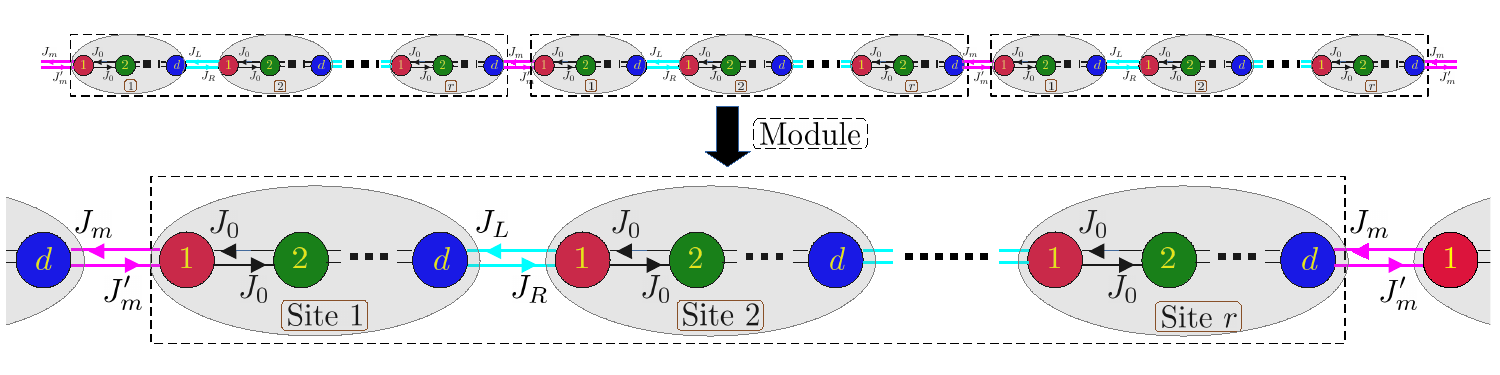} 
\caption{\textbf{Schematic of the model}. 
Each module consists of $r$ sites and each site has $d$ sublevels.
The modules, sites, and sublevels are denoted by dashed rectangles, grey ellipses, and colored circles, respectively.
The NH tunneling strengths between the modules are $J_m$ and $J'_m$ while those between sites are $J_L$ and $J_R$.
Coupling between adjacent sublevels is $J_0$ and is Hermitian.
The numbers of sublevels $d{=}1$ and $d{=}2$ correspond to modular Hatano-Nelson and NH SSH models, respectively.
}
\label{fig:schematic}
\end{figure*}

Discovery of topological phases in quantum systems is one of the crowning achievements of modern condensed matter physics~\cite{hasan2010colloquium, wen2017colloquium}.
From the perspective of quantum technologies, robustness of such topologically protected systems against local noises make them especially attractive~\cite{nayak2008non}.
In recent years, treatment of NH topological systems~\cite{lee2016anomalous, Shen2018Topological, Yao2018Edge} have attracted a lot of attention due to their role as effective description of open quantum systems~\cite{Ashida2020Non, Bergholtz2021Exceptional, Okuma2023Non}. 
The distinct topological features in NH physics emerge from the complex energy spectrum, with distinct forms of gap structures (such as point-gap and line-gap)~\cite{Gong2018Topological, Kawabata2019Symmetry}. 
NH systems have been experimentally realized in a wide variety of physical systems including electric circuits~\cite{Helbig2020Generalized, Hofmann2020Reciprocal, Liu2021Non}, acoustic~\cite{Zhang2021Acoustic, Gao2022Anomalous, pu2026acoustic} and photonic~\cite{Weidemann2020Topological, Song2020Two} lattices, mechanical metamaterials~\cite{Brandenbourger2019Non, Ghatak2020Observation}, lossy optical lattices~\cite{Lapp2019Engineering, Gou2020Tunable} and photonic quantum walkers~\cite{xiao2017observation, Xiao2020Non, wang2021detecting}.
A core feature of the NH topological systems is the exhibition of a sharp distinction between periodic boundary conditions (PBC) and open boundary conditions (OBC) persisting even in the thermodynamic limit. 
This leads to a discrepancy between the emergence of topological edge states in a finite system with the topological invariants calculated for bulk system with translational invariance.
Therefore, it is naturally intriguing to study the effect of an added periodicity of a modular structure to a NH topological system along with their possible applications in quantum technologies.

Quantum sensing, i.e., use of quantum properties to enhance the precision of estimation of parameters of interest beyond classically achievable precision limits, has now firmly emerged as a viable near-term quantum technology ~\cite{degen2017quantum, braun2018quantum, ye2024essay, ghosh2026journey}. 
Several resources for quantum enhanced sensitivity have been identified, including: (i) special form of entangled Greenberger-Horne-Zeilinger (GHZ) states~\cite{giovannetti2004quantum, giovannetti2006quantum, giovannetti2011advances}; (ii) squeezed optical~\cite{maccone2020squeezing, pezze2008mach, schnabel2017squeezed, polino2020photonic} or spin states~\cite{ma2011quantum, frerot2018quantum}; and (iii)  critical states which emerge at the boundary of different many-body phases~\cite{montenegro2025review}. 
Major theoretical progress have been made on sensing near different types of criticalities including first-order~\cite{raghunandan2018high, mirkhalaf2018supersensitive, sarkar2025first}, second-order~\cite{venuti2007quantum, zanardi2008quantum, gammelmark2011phase, skotiniotis2015quantum, rams2018limits, garbe2020critical, montenegro2021global, mukhopadhyay2025saturable}, Floquet~\cite{mishra2021driving}, time crystal~\cite{iemini2023floquet, montenegro2023quantum}, Stark~\cite{he2023stark} and quasi-periodic~\cite{sahoo2024localization} localization, and topological~\cite{sarkar2022free} phase transitions.
Recent advances in engineered quantum systems have also enabled their experimental realisation on various platforms~\cite{liu2021experimental, ding2022enhanced, beaulieu2025criticality, yu2025experimental, li2025nonequilibrium, wang2026quantum, liu2026enhanced}.
Multiplexing the number of critical points by considering a modular construction in the Hermitian setting adds further advantage to criticality-enhanced sensing~\cite{mukhopadhyay2024modular}.
The closing of the energy gap at criticality is now widely recognized as the key reason for the enhancement~\cite{abiuso2025fundamental}. 
While sensors based on NH systems have been widely studied for estimating boundary perturbations both theoretically~\cite{Wiersig2014Enhancing, Langbein2018No, Lau2018Fundamental, Zhang2019Quantum, Chen2019Sensitivity, budich2020non, Koch2022Quantum, Schomerus2020Nonreciprocal, McDonald2020Exponentially, Edvardsson2022Sensitivity, Ding2023Fundamental} and experimentally~\cite{Liu2016Metrology, Hodaei2017Enhanced, Yu2020Experimental, Wang2020Petermann, xiao2024non, Yu2024Heisenberg}, the NH gap closing mechanisms have also been explored recently for sensing bulk Hamiltonian parameters~\cite{sarkar2024critical, agarwal2025critical, xiao2026observation, agarwal2026quantum}. 
Therefore, it is crucial to study whether the modified gap structures in modular NH topological systems can be advantageous for sensing purposes.
Furthermore, as the modular structure introduces additional parameters in the system, it is equally important to investigate the capability of the NH sensors for multi-parameter estimation~\cite{liu2019quantum, ragy2016compatibility, carollo2019quantumness, albarelli2020perspective, pezze2025advances, mondal2025multicritical}.

In this work, we introduce a general 1D tight-binding NH model with modular structure and study the topological properties.
We then investigate the metrological advantages of such a system for criticality-enhanced sensing.
Our analysis uncovers several significant results which are summarized in the following.
(i) We present an analytical point-gap closing condition that induces the spectral topological phase transition for general modular couplings. 
This enables us to identify the critical points for enhanced sensitivity and also move them around by changing the modular structure.
(ii) We showcase the additional band topological phase transitions that occur due to the modular structures and quantify the phases by calculating the modified winding number in the NH domain.
(iii) We show the advantage of modular NH sensors over their non-modular counterparts as the significant enhancement of sensitivity gained by tuning the modular couplings in certain parameter regime.
This can be achieved both for single- and two-parameter sensing.
(iv) We also show critical-enhancement for three-parameter sensing that can only be achieved with the modular systems.

\section{Modular Non-Hermitian Topology}
\label{sec:modular}

\begin{figure*}
\centering
\includegraphics[width=0.87\linewidth]{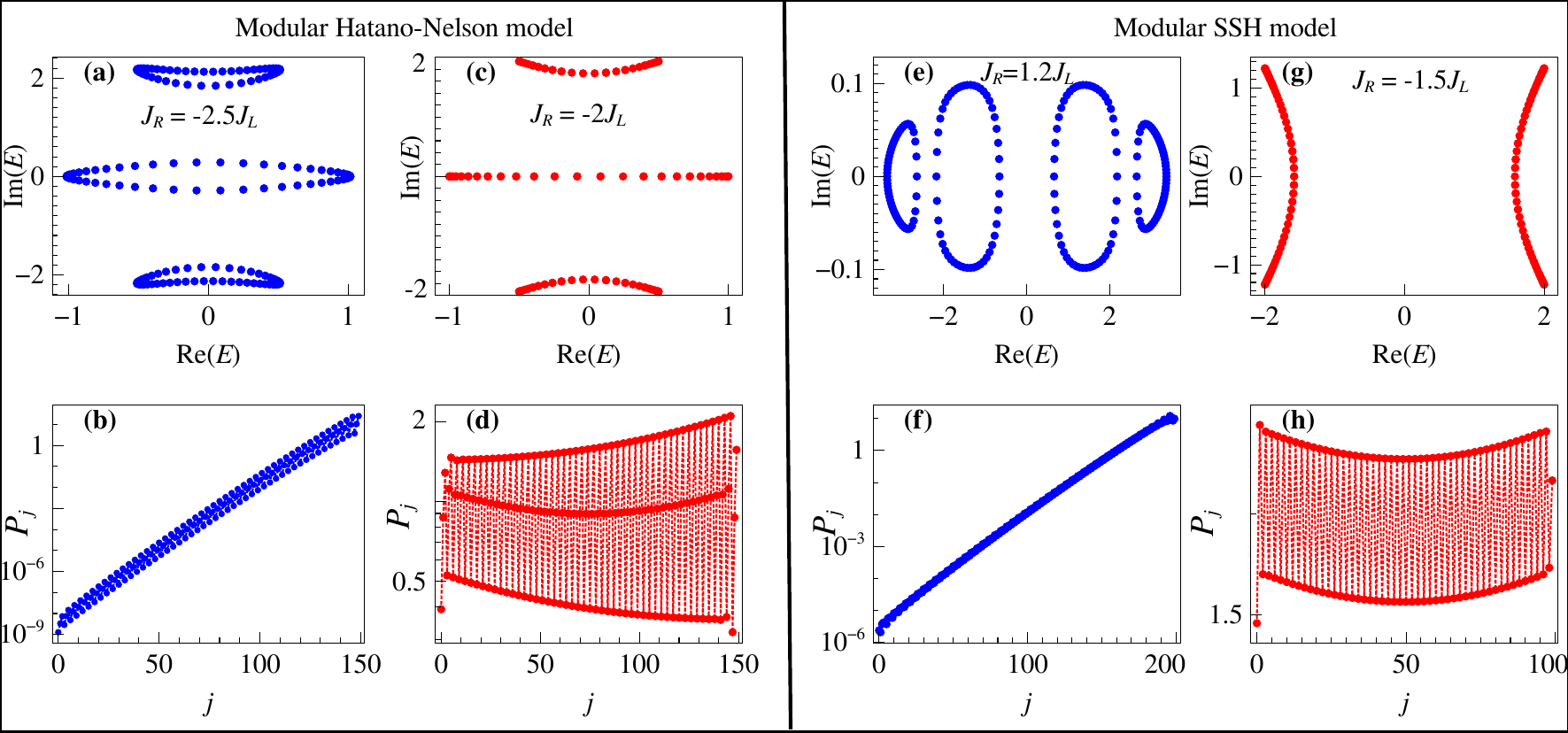} 
\caption{\textbf{Spectral topological phase transition and point gap structure}.
Left panel shows the case of modular Hatano-Nelson model with $d{=}1, r{=}3, L{=}50$. 
Here $J_m {=} J {=} 1/J'_m$ and $J{=}2J_L$.
(a) The point gap structure at $J_R{=}-2.5J_L$ is shown by three loops corresponding to the three bands of the model with PBC.
(b) The presence of point gap with PBC corresponds to NH skin effect for OBC, which is shown by the exponential edge localization of the cumulative population $P_j$ of all the eigenstates at site $j$.
(c) The closure of point gap at $J_R{=}-2J_L$.
(d) The corresponding vanishing of skin effect shown by the delocalization of the cumulative population.
Right panel shows the case of modular NH SSH model with $d{=}2, r{=}2, L{=}50, J_0{=}2J_L$. 
Here $J_m {=} J_L{+}J, J'_m {=} J_R{+}J$ and $J{=}0.5J_L$.
(e) The point gap structure at $J_R{=}1.2J_L$ is shown by four loops corresponding to the four bands of the model with PBC.
(f) The corresponding NH skin effect for OBC.
(g) The closure of point gap at $J_R{=}-2.5J_L$.
(h) The corresponding vanishing of skin effect.
}
\label{fig:point}
\end{figure*}

We start with an analysis of the topological properties of a general NH model in one-dimension with modular structure.
We particularly emphasize on the effect of the modular couplings on the topological phase transitions.
To do so, we first introduce the model and then focus separately on two types of topological phase transitions.
We first look at the topology of the spectrum, corresponding to an exclusively NH energy gap structure known as the point gap.
We then discuss the band topology of the system which is associated with a different energy gap structure known as the line gap.

\subsection{Model}
\label{sec:model}

We consider a modular version of a 1D tight-binding NH system comprising of $L$ modules, each of which has $r$ sites, and each site has $d$ sublevels.
The total number of sites is therefore $N {=} rL$.
The model is shown in Fig.~\ref{fig:schematic}.
The non-Hermiticity comes from non-reciprocal tunneling strengths between the sites ($J_L$ to the left, $J_R$ to the right) and between the modules ($J_m$ to the left, $J'_m$ to the right).
We assume Hermitian coupling $J_0$ between the adjacent sublevels.
We can write down the Hamiltonian as  
\begin{align}
    H = & \sum_{n=1}^{L-1} \, \sum_{\mu=1}^{r-1} \, \sum_{\nu=1}^{d-1} \Big[J_0 \ket{n, \mu, \nu} \bra{n, \mu, \nu{+}1} + \text{H.c.} \Big] \nn
    + & \Big[J_L \ket{n, \mu, d} \bra{n, \mu{+}1, 1} + J_R \ket{n, \mu{+}1, 1} \bra{n, \mu, d} \Big] \nn
    + & \Big[J_m \ket{n, r, d} \bra{n{+}1, 1, 1} + J'_m \ket{n{+}1, 1, 1} \bra{n, r, d} \Big],
    \label{eq:ham}
\end{align}
where $n, \mu, \nu$ denote the indices for the module, site, and sublevels, respectively.
Note that, this Hamiltonian is written with the OBC.
To invoke PBC, one can let the module index $n$ run from 1 to $L$, where $n {=} L{+}1$ is equivalent to $n {=} 1$.
For $d {=} 1$, this Hamiltonian corresponds to the modular version of Hatano-Nelson model~\cite{Hatano1996Localization}
\begin{align}
    H_{\rm HN} {=} \sum_{j} \Big(J_L \ket{j{-}1} \bra{j} {+} J_R \ket{j{+}1} \bra{j} \Big) .
    \label{eq:HN-ham}
\end{align} 
On the other hand, $d {=} 2$ corresponds to the modular NH version of the Su–Schrieffer–Heeger (SSH) model~\cite{Su1979Solitons, Yao2018Edge, Lieu2018Topological}
\begin{align}
    H_{\rm SSH} &= \sum_{j} \Big(J_0 \ket{j, A} \bra{j, B} + J_0 \ket{j, B} \bra{j, A} \nn
    &+ J_L \ket{j{-}1, B} \bra{j, A} + J_R \ket{j{+}1, A} \bra{j, B} \Big)
    \label{eq:SSH-ham}
\end{align}
with sublevels $A$ and $B$.

The complex spectrum of the NH Hamiltonian $H$ gives rise to rich band structures and topological properties~\cite{Gong2018Topological}.
The gap closing instances in the complex energy plane denote interesting topological phase transitions beyond those present in the Hermitian domain~\cite{Kawabata2019Symmetry}. 
In the following subsections, we show how the modular structure modifies the spectrum and introduces new instances of gap closing.

\subsection{Point gap and spectral topology}
\label{sec:point}

For a 1D NH topological system with PBC, the spectrum can create one or multiple loops in the complex energy plane as the quasi-momentum $k$ goes around the Brillouin zone.
In such cases, each energy value inside the loop constitutes a point gap~\cite{Gong2018Topological}.
Correspondingly, the eigenstates of the system with OBC are edge-localized and this is known as the NH skin effect~\cite{Yao2018Edge, Martinez2018Non}.
The edge localization can also be thought of as the result of modes with group-velocities directed towards that edge dominating in the long-time limit~\cite{gohsrich2025non}.
As the Hamiltonian parameters vary, it is possible to contract the loop structure into a curve, thus resulting in a point gap closure. 
This is known as a spectral topological phase transition which can be captured by a sudden change in the winding number of the spectrum~\cite{Gong2018Topological}.
Consequently, the eigenstates of a finite system with open boundary condition gets delocalized and the skin effect disappears~\cite{Borgnia2018Non, Okuma2020Topological, Zhang2020Correspondence}.
This critical phenomenon can be utilized to estimate the parameter that drives the transition~\cite{sarkar2024critical, xiao2026observation}.
We now derive the conditions for point gap closing and the application to metrology will be discussed in the next section.

A theoretical understanding of the edge localization of the eigenstates for OBC, due to the presence of point gap in the PBC spectrum, is given by the generalized Brillouin zone (GBZ) formalism~\cite{Yao2018Edge, Yokomizo2019Non}.
Here, the ansatz for an eigenstate of $H$ is written as $\ket{\psi} = \sum_{n,\mu,\nu} \beta^n c_{\mu, \nu} \ket{n, \mu, \nu}$ with $\beta$ denoting the localization parameter that determines the GBZ.
Now applying the eigen-equation $H \ket{\psi} {=} E \ket{\psi}$ to a module in the bulk leads to a quadratic equation $a \beta^2 + b \beta + c {=} 0$, with
$a {=} J_0^{r(d-1)} \, J_L^{r-1} \, J_m, c {=} J_0^{r(d-1)} \, J_R^{r-1} \, J'_m$.
See Appendix~\ref{app:spectral} for derivation.
Therefore the solutions $\beta_{1}$ and $\beta_{2}$ (for a given value of $E$) satisfy $\beta_{1} \beta_{2} {=} \frac{J_R^{r-1} \, J'_m}{J_L^{r-1} \, J_m}$.
The $\beta$ parameter serves as a generalization of the Bloch phase factor $e^{ik}$ in bulk hermitian systems~\cite{Yao2018Edge}.
Therefore, for NH systems, it establishes the equivalence of the continuum bulk band spectrum with the spectrum of the chain with OBC in the thermodynamic limit.   
This requires certain constraint on the values $\beta$ can take, which was derived in Ref.~\cite{Yokomizo2019Non} for general 1D NH systems. 
In our case it boils down to $|\beta_{1}| {=} |\beta_{2}|$.
On the other hand, the condition for point-gap closure demands the amplitude of the localization parameter $\beta$ to be one.
Therefore, it is given by $|\beta_{1}| {=} |\beta_{2}| {=} 1$ which occurs for
\begin{align}
    \left| \frac{J_R^{r-1} \, J'_m}{J_L^{r-1} \, J_m} \right| = 1
    \label{eq:pgc} .
\end{align}
The left hand side of Eq.~(\ref{eq:pgc}) is a quantifier of the overall non-reciprocity in tunneling. 
The numerator is the product of all the tunneling parameters in a module to the right, while the denominator is the product of those to the left.
Thus, the gap closing condition conveys the fact that when the overall non-reciprocity is absent, the skin effect vanishes and the phase transition occurs. 
This expression also implies that $J_R/J_L$ can take any complex value with amplitude $|(J_m/J'_m)^{\frac{1}{r-1}}|$ to induce the phase transition.
While in absence of the modular structure, the point gap closes at $|J_R/J_L| {=} 1$, the gap closing location can be shifted by varying $r$ and the forms of $J_m$ and $J'_m$ in the modular case.
For example, if the modular couplings are set by a single parameter $J$, and we take $J_m {=}J{=} 1/J'_m$, then the point gap closes at $|J_R/J_L| {=} |J^{\frac{2}{r-1}}|$.
On the other hand, if the modular couplings are taken to be just shifted values of inter-site couplings inside the module, i.e., $J_m {=} J_L{+}J, J'_m {=} J_R{+}J$, then it is easy to show from Eq.~\eqref{eq:pgc}, that the point gap closes at $J_R {=} -(J_L{+}J)$ for $r{=}2$.

\begin{figure*}
\centering
\includegraphics[width=0.95\linewidth]{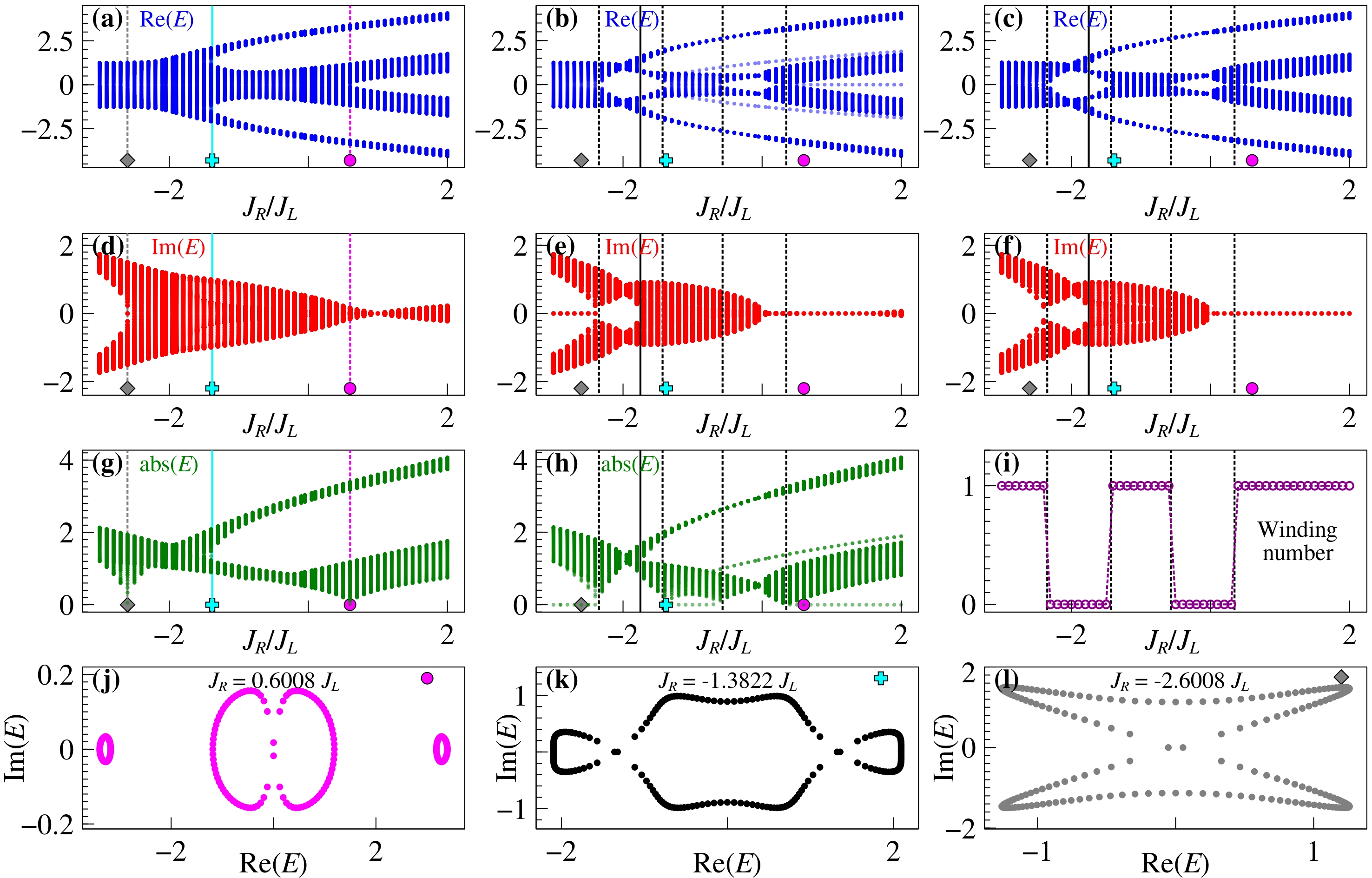} 
\caption{\textbf{Band topological phase transition and line gap structure}. 
The band formation in the spectrum corresponding to different boundary conditions are shown for $d{=}2, r{=}2, L{=}50, J_0{=}1.25J_L$. 
Here $J_m {=} J_L{+}J, J'_m {=} J_R{+}J$ and $J{=}2J_L$.
For band structure with PBC, (a), (d), (g) show real, imaginary, and absolute energies, respectively. 
The vertical lines denote $J_R$ values for different types of line gap closures. 
The magenta circle is for $J_R{=}0.6008J_L$. Here closure of the line gap between the two central bands are shown in (j).
The cyan cross is for $J_R{=}-1.3822J_L$. Here closure of the line gap between the central band and each of the side bands are shown in (k).
The grey triangle is for $J_R{=}-2.6008J_L$. Here the dissociation of the single band into two band is shown in (l). 
For band structure with OBC, (b), (e), (h) show real, imaginary, and absolute energies, respectively.
The (c) real and (f) imaginary part of the spectrum of the generalized Bloch Hamiltonian $H_{\beta}$.
The black dashed vertical lines show the gap closing points between the central bands in the spectrum with OBC and spectrum of $H_{\beta}$.
The black solid vertical lines show the gap closing points between the central and side bands.
(i) Winding number calculated with $H_{\beta}$.
}
\label{fig:line}
\end{figure*}

We showcase this result with two examples in Fig.~\ref{fig:point} where we consider real values of the couplings.
In this paper, we set $\hbar {=} 1$ and take $J_L{=}1$ to be the unit of the coupling strength, which fixes the time and energy units.
We first consider $d{=}1$ (modular Hatano-Nelson model) and look at the case with $J_m {=}J{=} 1/J'_m$ with $J{=}2J_L$.
In Fig.~\ref{fig:point}(a), the three bands are shown for $r{=}3$ as three loops for system with PBC at $J_R{=}-2.5J_L$.
The corresponding skin effect for OBC is shown in Fig.~\ref{fig:point}(b) with the exponential localization of the cumulative population at site $j$ as $P_j {=} \sum_m |c_{m,j}|^2$, where the $m$-th eigenstate of $H$ is given by $\ket{\psi_m} {=} \sum_{j=1, \dots, N} c_{m,j} \ket{j}$.
The point gap closure happens at $J_R{=}-2J_L$ which is a solution of Eq.~\eqref{eq:pgc}, as can be seen in Fig.~\ref{fig:point}(c).
The corresponding delocalization of the eigenstates is shown in Fig.~\ref{fig:point}(d).
We then consider the case of $d{=}2$ (modular SSH model) with $r{=}2$ and modular couplings $J_m {=} J_L{+}J, J'_m {=} J_R{+}J$.
As mentioned before, the criticality occurs at $J_R {=} -(J_L{+}J)$.
As we set $J{=}0.5J_L$, the point gap structure and the corresponding skin effect are depicted in Figs.~\ref{fig:point}(e) and (f), respectively for $J_R{=}1.2J_L$.
The point gap closure and vanishing of skin effect for $J_R{=}-1.5J_L$ are shown in Figs.~\ref{fig:point}(g) and (h), respectively.

\subsection{Line gap and band topology}
\label{sec:line}

While the spectral topology associated with the point gap is unique to the NH systems, the conventional band topology in the Hermitian systems gets extended in NH models with line gap structure~\cite{Kawabata2019Symmetry}.
A line gap occurs in the form of a reference line in the complex energy plane that separate different bands~\cite{Gong2018Topological}.
Band topological phase transitions are linked with line gap closure which can be detected by looking at the spectrum of the Bloch Hamiltonian $H_k$ associated with our model.
As $H_k$ has dimension $rd$, it is possible to analytically diagonalize it for small values of $rd$, beyond which one has to resort to numerical methods.
However, the line gap closing points thus found are slightly deviated from values at which the actual emergence of topological edge states occur in a finite system with OBC.
Such disagreement is a standout feature for NH systems with skin effect which localizes both the topological edge state and the bulk state in the non-topological phase~\cite{Yao2018Edge}. 
To remedy this breakdown of bulk-boundary correspondence in NH systems~\cite{Kunst2018Biorthogonal}, one can employ the GBZ formalism to construct a generalized Bloch Hamiltonian $H_{\beta}$ by simply replacing the $e^{ik}$ by $\beta$~\cite{Yao2018Edge, Yokomizo2019Non}.
The spectrum of $H_{\beta}$ gives the bulk bands corresponding to the spectrum with OBC in thermodynamic limit and therefore predicts the phase transitions correctly.

Without a modular structure, the band topology requires $d {\ge} 2$. 
For example, the non-modular NH SSH model (Eq.~\eqref{eq:SSH-ham}) has two bands with the gap closing occurring at $J_R {=} J_0$ or $J_L {=} J_0$ for PBC.
However, for the spectrum with OBC, the gap closes for $J_0^2 {=} \pm J_R J_L$~\cite{Yao2018Edge}.
This gap closing point is recovered by modifying the Bloch Hamiltonian as prescribed above.
In the presence of a modular structure, the number of bands increases $r$ times, giving rise to other band-gap closing possibilities.
Analogous to the Hermitian case~\cite{lee2022winding}, we anticipate one type of gap closing to occur between the two central bands and another type of simultaneous gap closing between all the adjacent bands.

Therefore, we compute the band structure for the modular SSH model ($d{=}2$) with $r{=}2$ and illustrate the results in Fig.~\ref{fig:line}.
With $J_m {=} J_L{+}J, J'_m {=} J_R{+}J$ and $J{=}2J_L$, the spectrum with PBC can be obtained by diagonalizing the Bloch Hamiltonian
\begin{align}
    H_k = 
    \begin{pmatrix} 
    0 & J_0 &  0  & J'_m e^{-ik} \\ 
    J_0 & 0 & J_L & 0 \\
    0 & J_R & 0 & J_0 \\
    J_m e^{ik} & 0 & J_0 & 0 \\
    \end{pmatrix} . 
    \label{eq:Hk}
\end{align}
Explicit expressions for gap closing conditions can be found in  Appendix~\ref{app:band}, and has been used in Figs.~\ref{fig:line}(a, d, g) where the line gap closing points are marked.
The vertical lines denote $J_R$ values for different types of line gap structures. 
The closure of the line gap between the two central bands are shown in Fig.~\ref{fig:line}(j) for $J_R{=}0.6008J_L$, which reduces the number of bands to three.
The closure of the line gap between the central band and each of the side bands are shown in Fig.~\ref{fig:line}(k) for $J_R{=}-1.3822J_L$, which results in formation of a single band.
The dissociation of the single band into two bands is shown Fig.~\ref{fig:line}(l) for $J_R{=}-2.6008J_L$, with emergence of line gap again. 
However, the spectrum with OBC shows different transition points due to the presence of skin effect, as shown in Figs.~\ref{fig:line}(b, e, h).
We observe that the gap closing between the central band and the side band do not lead to a topological phase transition.
Instead, there is an additional topologically non-trivial phase due to gap closure between the two central bands.
This are clearly marked by the emergence of mid-gap edge states, as shown in Fig.~\ref{fig:line}(h).
To determine the transition points correctly, we construct the GBZ as the trajectory of $\beta$ (Appendix~\ref{app:numerics}).
The numerically solved $\beta$ values match well with the analytical prediction $|\beta|^2 {=} \frac{J_R \, J'_m}{J_L \, J_m}$.
Now we solve for the spectrum of the generalized Bloch Hamiltonian $H_{\beta}$.
As shown in Appendix~\ref{app:band}, diagonalization of $H_{\beta}$ allows us to predict the modified gap closing points between the two central bands at $J_R{=}0.3468J_L, -0.5685J_L, -1.4315J_L, -2.3468J_L$.
The gap closing point between the central and side bands is also shifted to $J_R{=}-1.75J_L$.
The first type of gap closings are shown as black dashed vertical lines in Fig.~\ref{fig:line}, whereas the second type of gap closing is denoted by black solid vertical line.
As Figs.~\ref{fig:line}(c, f) shows, the phase transition points are correctly restored.

We then proceed to calculate the topological invariant of the system to concretely capture the different topological phases.
For our 1D system with chiral symmetry, this invariant is a winding number.
The chiral symmetry becomes evident with a basis transformation that results in an off-diagonal form of the generalized Bloch Hamiltonian,
\begin{align}
    \tilde{H}_{\beta} = B H_{\beta} B^{-1} =
    \begin{pmatrix} 
    0 & 0 &  J_0  & J'_m /\beta \\ 
    0 & 0 & J_R & J_0 \\
    J_0 & J_L & 0 & 0 \\
    J_m \beta & J_0 & 0 & 0 \\
    \end{pmatrix} 
    \equiv
    \begin{pmatrix} 
    \boldsymbol{0} &  h_{\beta}^{+} \\
    h_{\beta}^{-} & \boldsymbol{0} \\
    \end{pmatrix},
    \label{eq:Hboff}
\end{align}
where 
\begin{align}
    B =
    \begin{pmatrix} 
    1 & 0 & 0 & 0 \\ 
    0 & 0 & 1 & 0 \\
    0 & 1 & 0 & 0 \\
    0 & 0 & 0 & 1 \\
    \end{pmatrix} .
\end{align}
The unitary chiral symmetry operator is $S {=} \mathds{1}_2 \otimes \sigma^z$ which satisfies $S \tilde{H}_{\beta} S^{-1} {=} -\tilde{H}_{\beta}$, with $\mathds{1}_2, \sigma^z$ as $2 {\times} 2$ identity and Pauli $z$ operators, respectively.
The winding number can be defined as~\cite{chiu2016classification, asboth2013bulk, maffei2018topological, he2021non}
\begin{align}
    w = \frac{1}{2 \pi i} \oint_{C_{\beta}} d\beta \ \partial_{\beta} \log{[\det{(h_{\beta}^{+})}]} ,
    \label{eq:wn}
\end{align}
where the integration is carried out on the GBZ formed by the trajectory of $\beta$ denoted by ${C_{\beta}}$~\cite{Yao2018Edge}.
As shown in Fig.~\ref{fig:line}(i), the winding number successfully captures the topological phase transitions.

While the above analysis showcases the results for a specific choice of module size, internal level dimension, and modular coupling choices, it also establishes the richness of the NH topological properties that can be realized by modifying these parameters.

\section{Modular non-Hermitian sensor}
\label{sec:sensor}

We now utilize the sensitivity of the eigenstates on the Hamiltonian parameters near the phase transition points and establish the critical enhancement of the precision of estimation.
The line gap closing mechanism is not particularly useful for this purpose as the localized edge state transitions into a bulk state which is also localized due to skin effect.
Therefore, we numerically observe no effect of a line gap closing transition in the precision bounds.
However, the spectral topological transition associated with point gap closing actually gives rise to a localization-delocalization transition and the eigenstates can be used as probes for sensing the parameter driving the transition with criticality-enhanced precision~\cite{sarkar2024critical, xiao2026observation}. 
As for the choice of the probe state, we focus on the steady state of the Hamiltonian $H$ (i.e., the right eigenstate that has the largest imaginary eigenvalue) as this is the most physically relevant state obtained with long-time evolution.
In the following subsections, we first give an overview of the parameter estimation framework and then report the advantages provided by the modular NH systems for both single- and multi-parameter estimation scenarios.
We anticipate such advantages to stem from the spatial structure in the wave-function due to modularity.
In the localized state, each module features a local skin effect due to imbalance in $J_R$ and $J_L$, which is further enveloped by the skin effect induced by $J_m$ and $J'_m$.
Due to the delocalzation at the point gap closing transition, the wave-function can go through additional or less structural changes than in the non-modular case, depending on the relative strengths of $J_L, J_R, J_m$, and $J'_m$.
Therefore, we expect the probe state to be more sensitive for certain choices of the coupling strength.

\begin{figure*}
\centering
\includegraphics[width=0.9\linewidth]{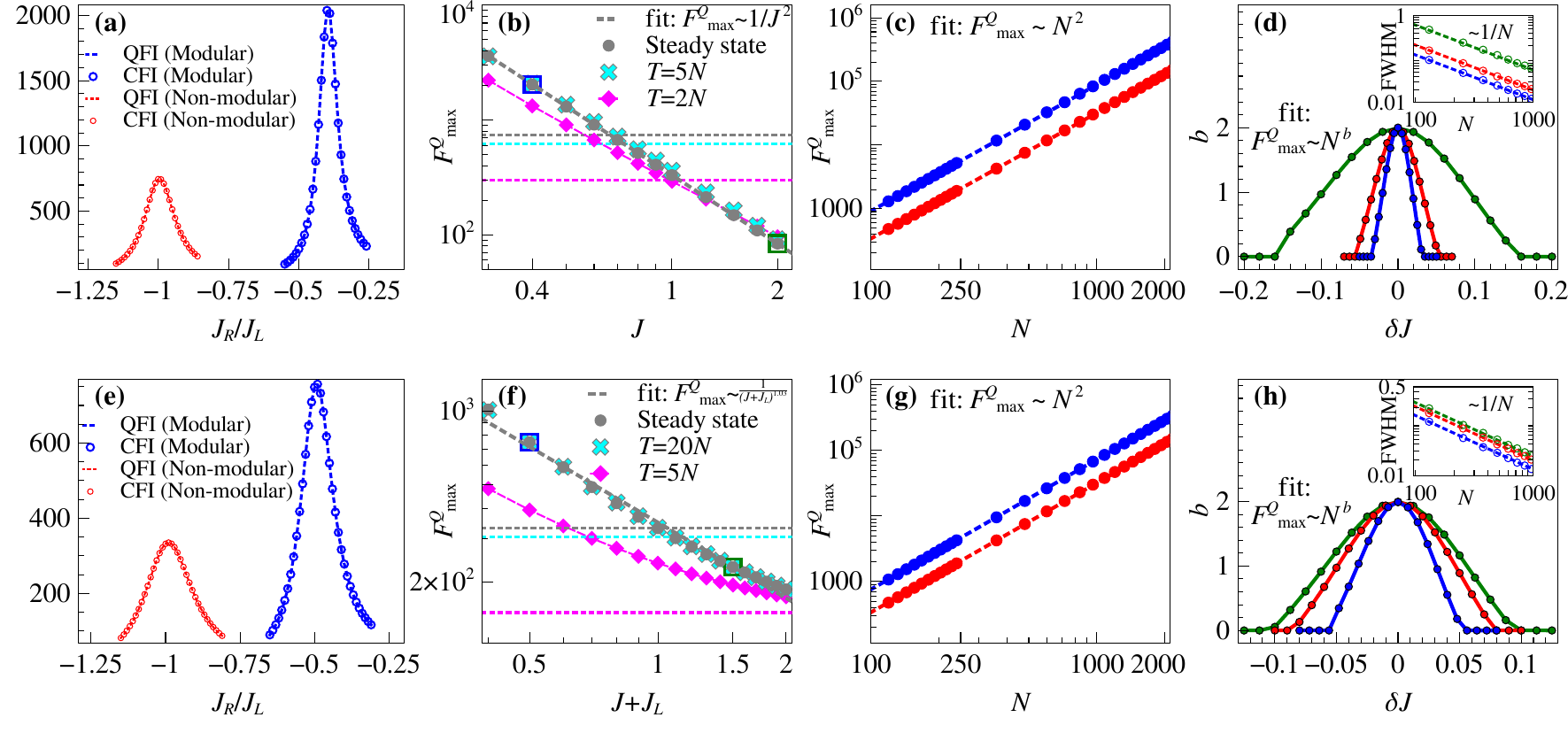} 
\caption{\textbf{Single parameter sensing}. 
Top panel correspond to modular Hatano-Nelson model and bottom panel correspond to modular SSH model.
(a, e) QFI and CFI (in the position basis) at the point gap closure. 
Blue curves correspond to the gap closing  for the modular case.
Red curves correspond to the non-modular case where the point gap closes at $J_R{=}-J_L$. 
For modular Hatano-Nelson model, gap closes at $J_R{=}-0.4J_L$ for $d{=}1, r{=}3, L{=}50, J_m {=} J {=} 1/J'_m$ and $J{=}0.4J_L$. 
For modular SSH model, gap closes at $J_R{=}-0.5J_L$ for $d{=}2, r{=}2, L{=}50, J_0{=}2J_L, J_m {=} J_L{+}J, J'_m {=} J_R{+}J$ and $J{=}-0.5J_L$.
(b, f) The peak QFI $F^Q_{\rm max}$ value (gray circles) for the modular case as a function of $J$. Gray dashed line shows the non-modular value. Cyan crosses and magenta diamonds show the QFI value obtained after a fixed probe preparation time $T$ while the cyan and magenta dashed lines show the corresponding non-modular QFI values.
(c, g) The blue and red curves show scaling of QFI with site number $N{=}r L$ for modular and non-modular cases, respectively.
(d, h) The decrease in the scaling exponent $b$ as $J_R/J_L$ is deviated from its critical value by $\delta J$.
Insets of (d, h) shows the full width at half maximum (FWHM) of the QFI peak structure as a function of $N$.
The blue and green curves in (d) and (h) correspond to $J$ values marked with blue and green squares in (b) and (f) respectively. The red curves correspond to the non-modular case. 
}
\label{fig:single}
\end{figure*}

\subsection{Parameter estimation methodology}
\label{sec:method}

For a general multi-parameter estimation scenario, the information about $l$ unknown parameters, $\boldsymbol{\theta} {=} (\theta_1, \theta_2,\ldots,\theta_l)$ is encoded in the probe state $\rho_{\boldsymbol{\theta}}$.
Measurement is performed with a positive operator-valued measure (POVM) $\{\Pi_s\}$ for which the probability of obtaining the $s$-th outcome is $p_s(\boldsymbol{\theta}) {=} \text{Tr}[\rho_{\boldsymbol{\theta}} \Pi_s]$.
An estimator is constructed to infer the values of $\boldsymbol{\theta}$ from the probability distribution.
The precision of such estimation can be quantified by the covariance matrix $[\text{Cov}({\boldsymbol{\theta}})]_{i,j} {=} \braket{\theta_i \theta_j} - \braket{\theta_i} \braket{\theta_j}$.
Here $\braket{x}$ is the statistical average for the parameter $x$ obtained from the measurement process repeated $M$ times.
For unbiased estimators, the lower bound for the precision of such estimation is given by the multi-parameter Cram\'{e}r-Rao inequality~\cite{liu2019quantum} $\text{Cov}({\boldsymbol{\theta}}) \geq \frac{1}{M} \mathcal{F}^{C \, -1}(\boldsymbol{\theta}) \geq \frac{1}{M} \mathcal{F}^{Q \, -1}(\boldsymbol{\theta})$.
Here $\mathcal{F}^{C}$ is the $l \times l$ basis-dependent classical Fisher information matrix (CFIM) with elements
\begin{align}
\mathcal{F}^{C}_{i,j} = \sum_{s} \frac{\partial_{\theta_i} p_s(\boldsymbol{\theta}) \, \partial_{\theta_j} p_s(\boldsymbol{\theta})}{p_s(\boldsymbol{\theta})} \,
\label{eq:CFIM} .
\end{align}
The positive semi-definite quantum Fisher information matrix (QFIM) $\mathcal{F}^{Q}$ gives a tighter bound that can be obtained by optimizing over all possible basis choices.
For a pure state $\ket{\psi_{\boldsymbol{\theta}}} $, QFIM can be written down as
\begin{align}
\mathcal{F}^{Q}_{i,j} = 4 \text{Re}(\braket{\partial_{\theta_i} \psi | \partial_{\theta_j} \psi} - \braket{\partial_{\theta_i} \psi | \psi} \braket{\psi | \partial_{\theta_j} \psi}) .
\label{eq:QFIM_pure}
\end{align}
To obtain some meaningful scalar inequalities from the Cram\'{e}r-Rao matrix inequality, one can use a positive weight matrix $W$ to write down $\text{Tr}(W \text{Cov}({\boldsymbol{\theta}})) \geq \text{Tr}(W \mathcal{F}^{-1})/M$.
For the choice of $W {=} \mathds{1}$, one gets the lower bound for the total uncertainty of estimation, given by the total variance $\sum_i \sigma^2_{\theta_i} {\geq} \text{Tr}(\mathcal{F}^{Q \, -1})/M$.
The single parameter case for the parameter $\theta {=} \theta_i$ can be obtained by choosing $W_{i,i} {=} 1$ while all the other elements are zero.
Straightforwardly, the single parameter Cram\'er-Rao inequality becomes $\sigma^2_{\theta} \ge 1/M F^C \ge 1/M F^Q$~\cite{braunstein1994statistical, paris2009quantum, liu2019quantum}.
Here, the classical Fisher information (CFI) $F^C$ is given by
\begin{align} 
    F^C = \sum_s \frac{(\partial_\theta p_s)^2}{p_s}
\label{eq:cfi} ,
\end{align}
and the quantum Fisher information (QFI) $F^Q$ is its upper bound. 
For a pure state $\rho_{\theta} = \ket{\psi_{\theta}} \bra{\psi_{\theta}}$ encoding the single parameter $\theta$, the QFI is given by
\begin{align} 
F^Q = 4\left(\braket{\partial_\theta \psi_{\theta}|\partial_\theta \psi_{\theta}} - |\braket{\partial_\theta \psi_{\theta}|\psi_{\theta}}|^2 \right)
\label{eq:qfi} .
\end{align}
In the NH domain, $\rho_\theta$ can be a right eigen-state of the NH Hamiltonian or can be obtained by a non-unitary evolution. 
Hence, the probe state needs to be normalized in order to produce normalized probability distributions upon measurements~\cite{Alipor2014Quantum, Yu2023Quantum, Xiao2020Non, Yu2024Heisenberg}.

\subsection{Single parameter sensing}
\label{sec:single}

We first look at the sensing capabilities of the modular NH system for enhancing the precision in a single parameter estimation problem.
Keeping all the other parameters fixed, estimation of $J_R/J_L$ is studied near the point gap closure.
We do this by calculating the QFI and CFI with respect to $J_R/J_L$ for the steady state of the Hamiltonian near the transition.
See Appendix~\ref{app:numerics} for further details.
We find that the position basis, given by $\Pi_j {=} \ket{j} \bra{j} \, (j \in [1, N])$, is the optimal basis.
Therefore, the CFI calculated in this basis saturates the QFI upper bound.
This helps us to quantify the advantages provided by the modular structure over a non-modular system that can be realized by setting $J_m {=} J_L, J'_m {=} J_R$.
We derive the scaling of the QFI from calculations on finite sized systems.
Note that, the transition point where the QFI profile peaks depends on the system size (see Appendix~\ref{app:numerics}).
However, such shifts in the transition point are not drastic and the peak QFI for a decent sized system is not far from the QFI at the analytical transition point.

We showcase the enhanced sensitivity features near the critical point in Fig.~\ref{fig:single}.
The enhanced QFI for the steady state is shown as a function of $J_R/J_L$ in Fig.~\ref{fig:single}(a) for the modular Hatano-Nelson model with $d{=}1, r{=}3, J_m {=} J {=} 1/J'_m$ with $J{=}0.4J_L$, where $J_R {=} -0.4J_L$ is a critical point.
Comparing the QFI features with that of the non-modular case clearly shows a significantly higher peak value $F^Q_{\rm max}$, suggesting a gain in precision.
As mentioned before, the CFI in the position basis also recreates this optimally.
In Fig.~\ref{fig:single}(b), we look at the behaviour of $F^Q_{\rm max}$ as a function of $J$ and benchmark it against the value obtained in the non-modular case.
We find that $F^Q_{\rm max}$ falls off as $1/J^2$ in a broad range, suggesting that lower $J$ values result in better sensitivity.
While this shows the sensing advantage of modular systems in certain parameter range, one should also take into account the probe preparation time $T$ for a better comparison with the non-modular case in a physical realization.
Therefore, we additionally compare the QFI of the state obtained after a fixed evolution time $T$, starting from a state localized at the center of the lattice.
We find $T{=}5N$ almost recreates the steady state behavior and therefore we also look at a shorter time $T{=}2N$ which may be experimentally more accessible.
We observe the sensing advantage with modular structure to be retained although the probe states have not reached the theoretically required steady state.
The critical enhancement of the QFI is evident from the quadratic scaling of $F^Q_{\rm max}$ with system size $N$, i.e., $F^Q_{\rm max} {\sim} N^b$ with $b {\approx} 2$, as shown in Fig.~\ref{fig:single}(c).
This enhancement diminishes as one moves away from the critical point.
In Fig.~\ref{fig:single}(d), we show this by plotting the exponent $b$ as a function of the distance $\delta J$ between $J_R/J_L$ and its critical value.
It further displays that by increasing $J$, one can broaden the range of $J_R/J_L$ in which favorable scaling of QFI can be obtained.
The downside is, one has to then trade this advantage with lower values of $F^Q_{\rm max}$.
On the other hand, the range in which enhanced sensing is obtained decreases with increase in $N$, as the QFI peak structure gets narrower and higher.
We plot the full width at half maximum (FWHM) with $N$ in the inset of Fig.~\ref{fig:single}(d), which shows that FWHM ${\sim} 1/N$.
We also find that the QFI features primarily depend on the forms of the modular couplings $J_m$ and $J'_m$ and not on the number of sublevels $d$.

\begin{figure*}
\centering
\includegraphics[width=0.9\linewidth]{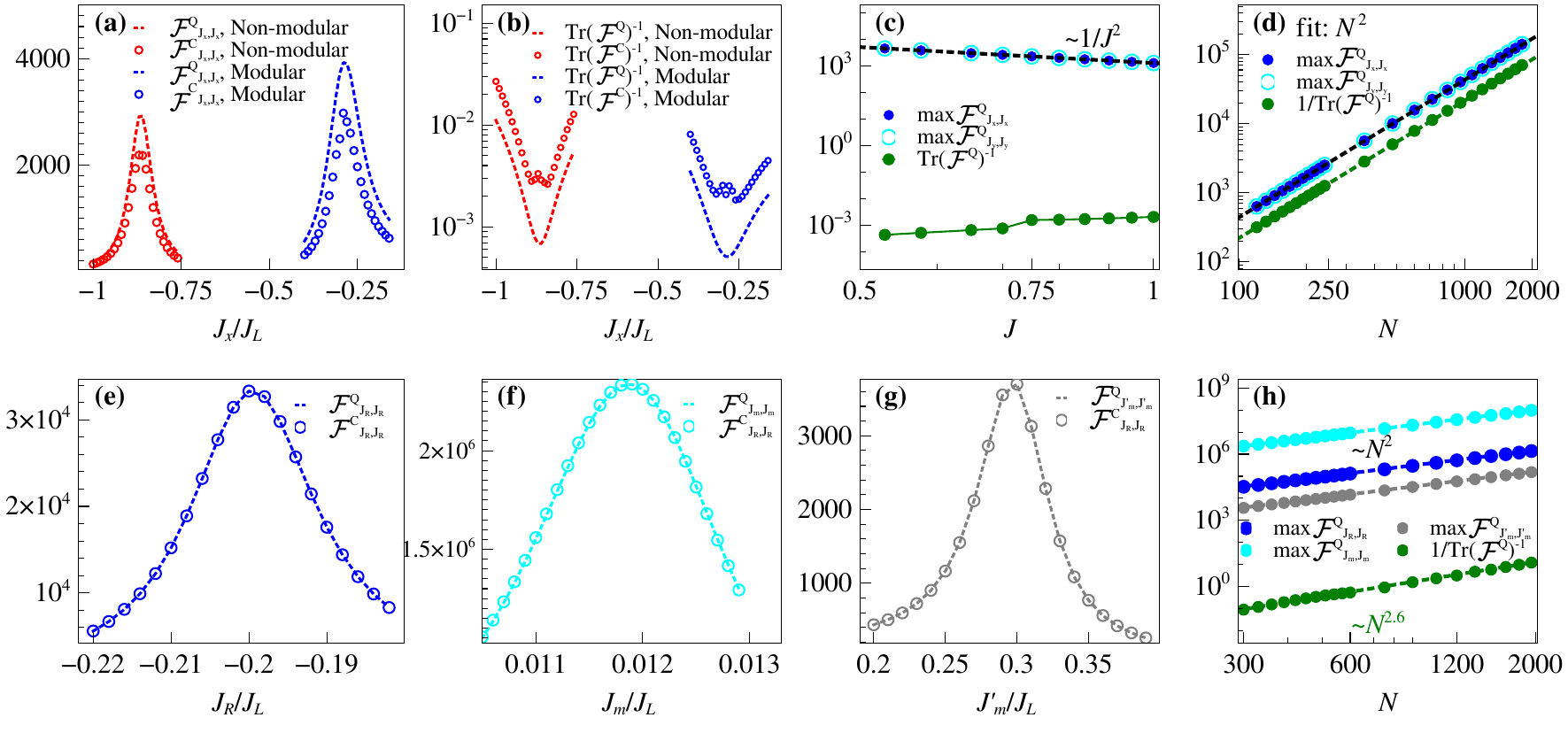} 
\caption{\textbf{Multi-parameter sensing}. 
Here we consider $d{=}1, r{=}3, L{=}100, J_m {=} J {=} 1/J'_m$.
Considering $J_R {=} J_x {+} i J_y$, we look at the QFIM $\mathcal{F}^{Q}(J_x, J_y)$ in the top panel. Blue and red curves show modular and non-modular cases, respectively.
(a) $\mathcal{F}^{Q}_{J_x, J_x}$ with $\mathcal{F}^{C}_{J_x, J_x}$ in the current operator basis. 
(b) $\text{Tr}(\mathcal{F}^{Q \, -1})$ with $\text{Tr}(\mathcal{F}^{C \, -1})$ in the current operator basis.
(c) $\mathcal{F}^{Q}_{J_x, J_x}$, $\mathcal{F}^{Q}_{J_y, J_y}$, and $\text{Tr}(\mathcal{F}^{Q \, -1})$ as functions of $J$.
(d) Quadratic scaling for $\mathcal{F}^{Q}_{J_x, J_x}$, $\mathcal{F}^{Q}_{J_y, J_y}$, $1/\text{Tr}(\mathcal{F}^{Q \, -1})$.
Bottom panel shows the modular multi-parameter sensing performance for estimating $(J_R, J_m, J'_m)$ near a critical value of (-0.2, 0.012, 0.3) given by Eq.~\eqref{eq:pgc}.
(e) $\mathcal{F}^{Q}_{J_R, J_R}$ with CFIM in the position operator basis.
(f) $\mathcal{F}^{Q}_{J_m, J_m}$ with CFIM in the position operator basis.
(g) $\mathcal{F}^{Q}_{J'_m, J'_m}$ with CFIM in the position operator basis.
(h) Quadratic scaling for $\mathcal{F}^{Q}_{J_R, J_R}$, $\mathcal{F}^{Q}_{J_m, J_m}, \mathcal{F}^{Q}_{J'_m, J'_m}$, and $N^{2.6}$ scaling for $1/\text{Tr}(\mathcal{F}^{Q \, -1})$.
}
\label{fig:multi}
\end{figure*}

The lower panel of Fig.~\ref{fig:single} shows similar behaviour for the modular SSH model with $d{=}2$, $r{=}2$, $J_0{=}2J_L$, $J_m {=} J_L{+}J, J'_m {=} J_R{+}J$ with $J{=}-0.5J_L$, where the transition occurs at $J_R {=} -(J_L+J) {=} -0.5J_L$.
The higher peak value of QFI and CFI over the non-modular case is shown in Fig.~\ref{fig:single}(e).
We find that the steady state in this case is reached after a longer evolution time $T{=}20N$.
A shorter preparation time $T{=}5N$ also shows the sensing advantage over the non-modular case.
In this case, we find that $F^Q_{\rm max}$ approximately goes as $1/(J+J_L)$, as shown in Fig.~\ref{fig:single}(f).
The quadratic scaling of $F^Q_{\rm max}$ is shown in Fig.~\ref{fig:single}(g).
As Fig.~\ref{fig:single}(h) shows, we again observe similar broadening of the range with enhanced sensitivity by increasing $J$ and $1/N$ dependence for the FWHM (inset).

\subsection{Multi-parameter sensing}
\label{sec:multi}

In the single parameter estimation case, we considered only real values of $J_R/J_L$.
However, the criticality condition in Eq.~\eqref{eq:pgc} suggests that by allowing $J_R/J_L {=} J_x {+} iJ_y$, one can use the NH system as a two-parameter sensor.
We find that it is indeed possible to achieve enhanced precision simultaneously for both $J_x, J_y$ as the criticality condition is set by $J_x^2 {+} J_y^2$.
We therefore construct the QFIM $\mathcal{F}^{Q}(J_x, J_y)$ and CFIM $\mathcal{F}^{C}(J_x, J_y)$ to study the scaling properties.
However, while the CFIM in the position basis also shows the critical enhancement, it turns out to be considerably sub-optimal.
For the case in hand with complex coupling strength, we take the measurement operators to be projectors formed by the eigenstates of the total particle current operator
\begin{align}
    \mathcal{I} = i \sum_{n=1}^{L-1} \, \sum_{\mu=1}^{r-1} \, \sum_{\nu=1}^{d-1} & \Big[\ket{n, \mu, \nu} \bra{n, \mu, \nu{+}1} - \ket{n, \mu, \nu{+}1} \bra{n, \mu, \nu} \Big] \nn
    + & \Big[\ket{n, \mu, d} \bra{n, \mu{+}1, 1} - \ket{n, \mu{+}1, 1} \bra{n, \mu, d} \Big] \nn
    + & \Big[\ket{n, r, d} \bra{n{+}1, 1, 1} - \ket{n{+}1, 1, 1} \bra{n, r, d} \Big]. 
\label{eq:current}
\end{align}
The results are shown in the top panel of Fig.~\ref{fig:multi}.
Here, we consider the case with $d{=}1, r{=}3, J_m {=} J {=} 1/J'_m$ and $J{=}J_L/\sqrt{3}$ to showcase the critical enhancement in the multi-parameter regime.
We show the results around the critical point $(J_x, J_y){=}(-J_L/\sqrt{12}, J_L/2)$.
The enhancement of the QFIM element $\mathcal{F}^{Q}_{J_x, J_x}$ is shown in Fig.~\ref{fig:multi}(a).
The CFIM calculated in the current operator basis shows that this basis is still sub-optimal, but produces values close to the QFIM elements.
Here we again observe the peak heights to be higher than those in the non-modular case.
The $\mathcal{F}^{Q}_{J_y, J_y}$ are not shown as they are the same as $\mathcal{F}^{Q}_{J_x, J_x}$.
In Fig.~\ref{fig:multi}(b), we plot $\text{Tr}(\mathcal{F}^{Q \, -1})$ that is associated with the total variance, and it again establishes the superior performance of the modular structure.
As shown in Fig.~\ref{fig:multi}(c), the peak values of $\mathcal{F}^{Q}_{J_x, J_x}$ and $\mathcal{F}^{Q}_{J_y, J_y}$ falls of quadratically with $J$ while $\text{Tr}(\mathcal{F}^{Q \, -1})$ increases with $J$.
This again suggests that modular NH sensors perform better with lower $J$ values.
Another crucial feature to consider in the present case is the attainability of the multi-parameter Cram\'{e}r-Rao bound which depends on the compatibility of the optimal measurement bases for the different parameters.
This attainability can be ensured by the weak compatibility condition $\text{Tr}(\rho [L_1, L_2]) {=} 0$~\cite{ragy2016compatibility}, where $L_j$ is the symmetric logarithmic derivative (SLD) for the $j$-th parameter~\cite{paris2009quantum}.
We numerically check that this condition is fulfilled near the transition point which shows that the ultimate bound can in principle be achieved simultaneously for both parameters.
We finally show the scaling of the critical value of $\mathcal{F}^{Q}_{J_x, J_x}$ , $\mathcal{F}^{Q}_{J_y, J_y}$ , and $1/\text{Tr}(\mathcal{F}^{Q \, -1})$ in Fig.~\ref{fig:multi}(d).
The quadratic scaling obtained for all these quantities establish the criticality-enhanced sensing capacity for estimating $J_x, J_y$ both individually and simultaneously.

As a final application in the multi-parameter scenario, we consider a case that is only achievable with the modular structure.
The criticality condition in Eq.~\eqref{eq:pgc} allows us to estimate the three parameters $J_R, J_m$, and $J'_m$ (while $J_L {=} 1$).
We show the three-parameter sensing capability in the lower panel of Fig.~\ref{fig:multi}.
The critical enhancement for $\mathcal{F}^{Q}_{J_R, J_R}$, $\mathcal{F}^{Q}_{J_m, J_m}$, and $\mathcal{F}^{Q}_{J'_m, J'_m}$ near a critical value of $(J_R, J_m, J'_m) {=} (-0.2, 0.012, 0.3)$ are shown in Figs.~\ref{fig:multi}(e), (f), and (g), respectively.
In this case the position basis again turns out to be the optimal basis for all the three parameters.
Since the compatibility condition is therefore satisfied trivially, we can deduce that the multi-parameter Cram\'{e}r-Rao bound can be attained in this case as well.
Hence, the corresponding CFIM elements match the QFIM values.
As displayed in Fig.~\ref{fig:multi}(h), quadratic scaling for all the diagonal elements of QFIM and even stronger scaling for $1/\text{Tr}(\mathcal{F}^{Q \, -1})$ show the criticality-enhanced capability for individual and joint estimation of all three parameters.

\begin{figure*}
\centering
\includegraphics[width=0.9\linewidth]{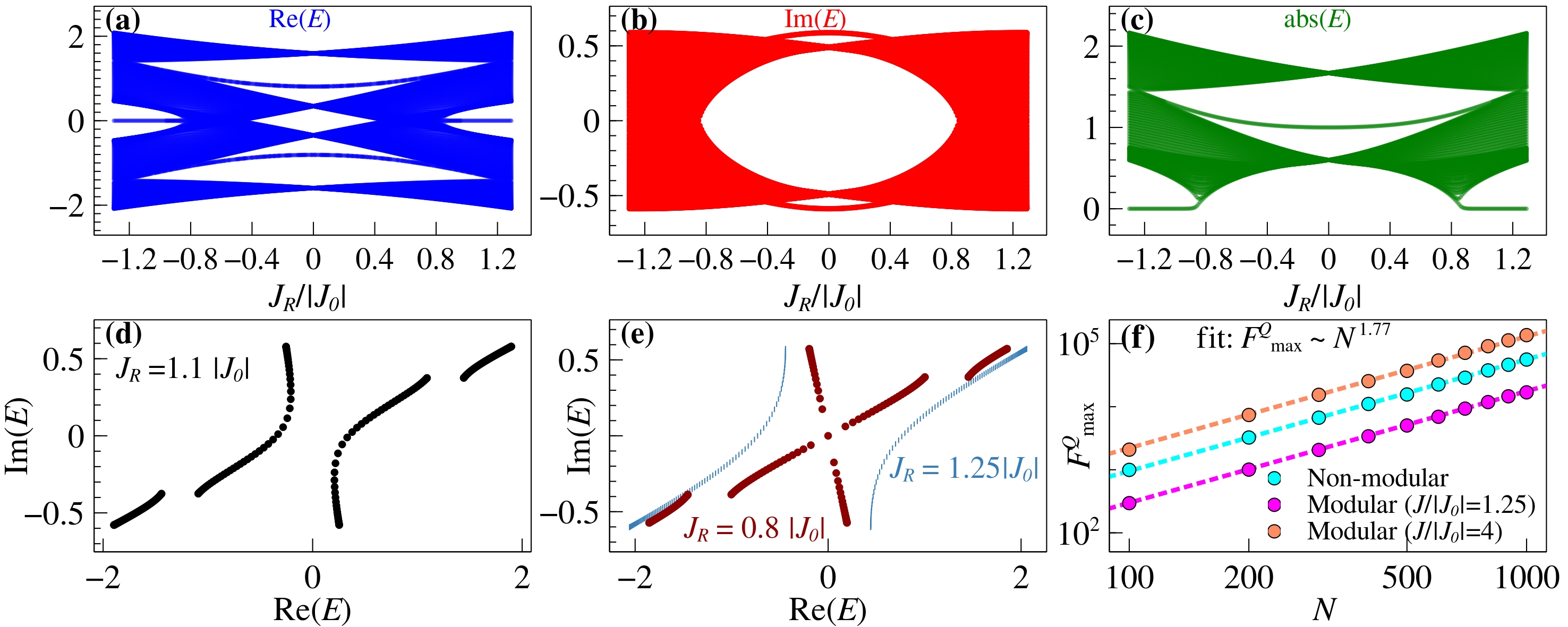} 
\caption{\textbf{Line gap without point gap}. 
Here we consider a modular NH SSH model with chiral and inversion symmetry with $d{=}2, r{=}2, J_0 {=} |J_0| e^{i \pi/5}$, $J_L {=} J_R, J_m {=} J {=} J'_m$, with $|J_0|$ as the unit.
For band structure with OBC for $J/|J_0| {=} 1.25$, (a), (b), (c) show real, imaginary, and absolute energies, respectively. 
(d) The four bands in the PBC spectrum for $J_R/|J_0| {=} 1.1$. 
(e) Closure of the line gap in the PBC spectrum between the two central bands at $J_R/|J_0| {=} 1/J {=} 0.8$ and between the central and side bands at for $J_R/|J_0| {=} J {=} 1.25$.
(f) Scaling of QFI peak $F^Q_{\rm max}$ with system size $N$ for the non-modular case ($J {=} J_R$) and modular cases with $J/|J_0| {=} 1.25$ and $J/|J_0| {=} 4$.
}
\label{fig:chiral}
\end{figure*}

\subsection{Sensing near line gap closing}
\label{sec:chiral}

Although line gap closing in the presence of a point gap is not metrologically useful, one can anticipate sensing advantage if the NH model does not feature point gap structure.
We examine this with an example based on a NH SSH model with chiral and inversion symmetries that requires the Bloch Hamiltonian to satisfy $\sigma^z H_k \sigma^z {=} -H_k$ and $\sigma^x H_k \sigma^x {=} H_{-k}$, respectively~\cite{Lieu2018Topological}.
A modular structure for such a model, based on our notations used in Eq.~\eqref{eq:ham} (or, in Fig.~\ref{fig:schematic}), is given by choosing $d{=}2$, $r{=}2$, $J_0 {=} |J_0| e^{i \phi}$, $J_L {=} J_R$, and $J_m {=} J'_m {=} J$, where we consider real $J_R$ and $J$.
Note that $|J_0|$ serves as the unit of energy in this subsection.
For the non-modular case given by $J {=} J_R$, the line gap closes at $J_R {=} \pm |J_0|$ for PBC and the zero-energy edge states appear for OBC when $|J_R/J_0| {>} 1$~\cite{Lieu2018Topological}.
The modular structure allows for four bands where the line gap between the two central bands close at $J_R/|J_0| {=} \pm 1/J$ and the line gap between a central band and the adjacent side band closes at $J_R/|J_0| {=} \pm J$ (see Appendix~\ref{app:band}).
The first kind of gap closing is associated with the emergence of zero energy edge states for OBC while the second kind of gap closing is connected to finite energy edge states.
As the zero energy state is numerically tractable, we consider this as the probe and calculate the QFI to estimate $J_R/|J_0|$.
The QFI peaks near the transition and we observe the peak value $F^Q_{\rm max}$ to be increasing with $J$.
Therefore, to surpass the sensing advantage obtained with the non-modular case, one needs to choose large enough $J$.
These findings are shown in Fig.~\ref{fig:chiral}.
The top panel shows the spectrum for the OBC case as a function of $J_R/|J_0|$ for $\phi {=} \pi/5$ and $J/|J_0| {=} 1.25$, which shows the two types of gap closings and the emergence of the associated edge states.
The four bands distinctly separated from each other at $J_R/|J_0| {=} 1.1$ are shown in Fig.~\ref{fig:chiral}(d).
The line gap closings between the two central bands at $J_R/|J_0| {=} 1/J {=} 0.8$ and between the central and side bands at $J_R/|J_0| {=} J {=} 1.25$ are shown in Fig.~\ref{fig:chiral}(e).
The scaling of $F^Q_{\rm max}$ with $N$ is shown in Fig.~\ref{fig:chiral}(f) where lower values such as $J/|J_0| {=} 1.25$ produces QFI lower than the non-modular case to surpass which one needs to consider higher values such as $J/|J_0| {=} 4$.
However, in all cases, the scaling exponent was found to be around $1.77$ with uncertainties of the order of $10^{-4}$.
We again observe the position basis to be close to an optimal basis in this scenario.

\begin{table*}
\centering

\begin{tabular}{|c|c|c|c|c|c|c|}
\cline{2-7}
\multicolumn{1}{c}{} & \multicolumn{2}{|c|}{Single-parameter sensing} & \multicolumn{2}{|c|}{Two-parameter sensing} & \multicolumn{2}{|c|}{Three-parameter sensing} \\
\cline{2-7}
\multicolumn{1}{c}{} & \multicolumn{1}{|c|}{$F^Q$ scaling} & Measurement basis & $1/\text{Tr}(\mathcal{F}^{Q \, -1})$ scaling & Measurement basis & $1/\text{Tr}(\mathcal{F}^{Q \, -1})$ scaling & Measurement basis \\
\hline
\shortstack{$J_m {=} J_L, J'_m {=} J_R$ \\ (Non-modular)}  & ${\sim} N^2$ & \shortstack{Position basis \\ (almost optimal)} & ${\sim} N^2$ & \shortstack{Current operator basis \\ (sub-optimal)} & - & - \\
\hline
$J_m {=} J {=} 1/J'_m$ & ${\sim} N^2$ & \shortstack{Position basis \\ (almost optimal)} & ${\sim} N^2$ & \shortstack{Current operator basis \\ (sub-optimal)} & - & - \\
\hline
$J_m {=} J_L{+}J, J'_m{=}J_R{+}J$ & ${\sim} N^2$ & \shortstack{Position basis \\ (almost optimal)} & - & - & - & - \\
\hline
$J_m, J'_m$ independent& - & - & - & - & ${\sim} N^{2.6}$ & \shortstack{Position basis \\ (almost optimal)} \\
\hline

\end{tabular}
\caption{Summary of various sensing scenarios near point gap closings considered in this work.}
\label{tab:summary}
\end{table*}

\subsection{Physical implementation}
\label{sec:exp}

\begin{figure}[b]
\centering
\includegraphics[width=\linewidth]{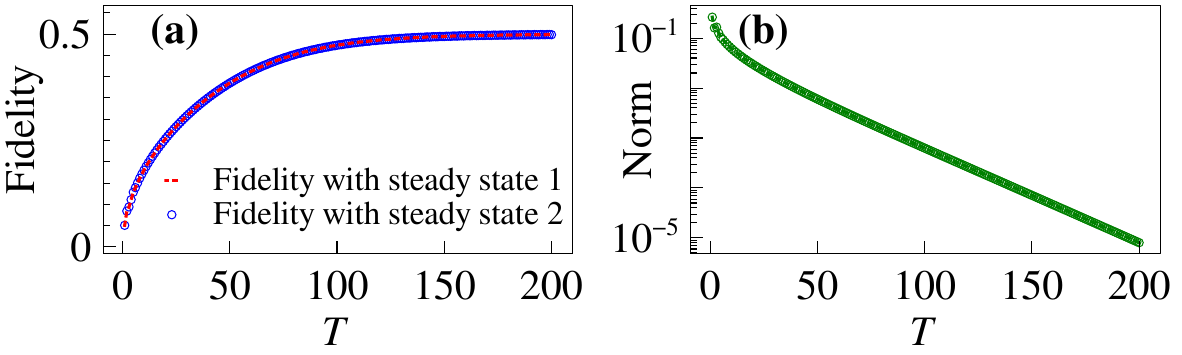} 
\caption{\textbf{Implementation details}. 
We consider discrete-time evolution of a state, initially localized at the lattice center, under modular Hatano-Nelson Hamiltonian with $d{=}1, r{=}2, L{=}30, J_m {=} J {=} 1/J'_m$ where the point gap asymptotically closes at $J_R{=}-0.25 J_L$ for $J{=}0.5 J_L$.
For the finite size considered, the actual gap closing occurs at $J_R{\approx}-0.22 J_L$.
(a) The fidelity of the evolved state with the two steady states under normalized conditions as a function of the discrete time step $T$.
(b) The norm of the state $\ket{\psi_T}$ evolved with the Hamiltonian along with a loss factor $e^{-\gamma}$ that appears in the discrete-time quantum walk implementation.
}
\label{fig:exp}
\end{figure}

While lossy optical lattices~\cite{Lapp2019Engineering, Gou2020Tunable} are viable candidates for implementing NH evolution required for probe preparation, significant modifications are needed to incorporate the modular structure in the lattices.
Therefore, we anticipate discrete-time non-unitary photonic quantum walk experiments~\cite{xiao2017observation, Xiao2020Non, wang2021detecting} to provide a suitable platform for this task where the sensing advantage near a point gap closing has already been observed~\cite{xiao2026observation}.
The chiral symmetry of the modular systems considered in this paper can be easily implemented in these quantum walk set-ups~\cite{mochizuki2020bulk, Xiao2020Non} by choosing suitable sequence of shift and coin operators along with an onsite gain-loss operator.
The shift operator $S$ moves the photon by one site towards left or right depending on its internal state and is realized with a beam displacer.
The coin operator $R_{\theta}$ creates rotation in the internal Hilbert space of the photon, typically in the form of $e^{-i \theta \sigma_y}$, and is implemented with half-wave plates.
The non-unitarity is introduced by a gain-loss operator $e^{\gamma \sigma_z}$ and can be realized with a partially polarized beam splitter that probabilistically reflects one of the internal states and transmits the other.
As it is not possible to coherently amplify the photon numbers, the action of non-unitary evolution operator $U$ is realized by evolution with $e^{-\gamma} U$ and subsequent renormalization at each step.
Essentially $U {=} e^{-i H}$, and $\gamma$ needs to be bigger than the maximum value of the imaginary energy of the steady state to avoid nonphysical amplification.
Note that the maximum imaginary energy is positive due to chiral symmetry.
In our case, the exact mapping of the modular couplings to the rotation angles in the coin operator is a non-trivial problem and we leave it for a future work.
We rather assume a valid mapping and investigate the feasibility of observing the sensing advantage in terms of the probe preparation time and analysis of successful measurement due to post-selection.
We take the case of modular Hatano-Nelson model with $r {=} 2, J_m {=} J {=} 1/J'_m$ and by choosing $J {=} 0.5$, look at the sensing performance near the point gap closing at $J_R {=} -J^2 {=} -0.25 J_L$.
For an experimentally relevant system with module number $L {=} 30$, the actual gap closing occurs at $J_R {\approx} -0.22 J_L$.
We take $\gamma {=} 0.49$ to surpass the maximum imaginary energy value of $0.468$.
There are two steady states where the real parts of the energy differ in sign. 
In Fig.~\ref{fig:exp}(a), we show the fidelity of the evolved state with the steady states and see that it reaches the steady value after about $T {=} 3N$, which is comparable with a similar system considered in Fig.~\ref{fig:single}(b).
The effect of photon loss is captured by the norm of the evolved state $\ket{\psi_T} {=} (e^{-\gamma} e^{-iH})^T \ket{\psi_0}$ where the initial state $\ket{\psi_0}$ is taken to be localized at the center of the lattice.
As shown in Fig.~\ref{fig:exp}(b), the norm decreases to $10^{-5}$ when the steady state is reached.
However, the merit of the quantum walk experiments is that number of photons injected in the beginning of the experiment can be in the range $10^4 - 10^6$, which is obtained by attenuating a pulsed laser beam with neutral density filters~\cite{xiao2017observation, xiao2026observation}.
This high number ensures successful detection even after such norm losses and the position measurement (which is close to an optimal basis) can be done with avalanche photodiodes.

\section{Conclusion}
\label{sec:conclusion}

Considering a general non-Hermitian tight-binding model in 1D, we study the effect of incorporating a modular structure on the topological properties.
The topological phase transitions are then leveraged for criticality-enhanced sensing.
For the spectral topology that has direct correspondence to the point gap structure, we analytically derive the phase transition condition for general modular couplings.
This enables us to shift the critical point of a non-modular system by changing the modular couplings, albeit within specified range allowed by the actual physical implementation platform.
Due to the localization-delocalization transition of the eigenstates at such criticalities, they are a viable resource for enhanced sensing.
On the other hand, the band topology is linked to the line gap structure in the spectrum, albeit loosely, due to the breakdown of conventional bulk-boundary correspondence in such NH systems featuring point gaps as well.
We show that the modular structure can introduce more instances of band-topological phase transitions and they can be captured by calculating the winding number in a generalized Brillouin zone instead of the conventional Bloch one.
In contrast to the spectral topological transitions, these criticalities do not change the localized feature of the eigenstates due to skin effect, and therefore, are not particularly useful for sensing.
In a single-parameter sensing scenario of a particular coupling strength that drives the spectral topological phase transition, the sensitivity shows quadratic scaling with system size.
For complex coupling parameter, this can be achieved effectively at any point on a circle in the complex plane as the criticality only depends on the amplitude of the parameter.
Various choices of modular couplings can increase the precision significantly compared to the non-modular case in suitable ranges of parameters.
Outside these ranges, the precision falls below the non-modular one but still show super-linear scaling.
Alternatively, the modular structure can also help in increasing the range of parameters around the transition point in which enhanced sensitivity can be achieved.
This applies for a two-parameter sensing case for estimating the real and complex part of the coupling simultaneously.
After establishing such advantages for single-parameter and two-parameter sensing, we also study a three-parameter sensing scenario with critical enhancement that can only be achieved in the modular case.
A summary of these results obtained for the different scenarios considered in this work are presented in Table~\ref{tab:summary}.
Our work thus paves the way for designing highly sensitive, reconfigurable sensors that exploit engineered criticality, and opens up new directions for multi-parameter estimation using non-Hermitian topological phase transitions beyond conventional sensing schemes.
Furthermore, it would be interesting to study the effect of modular structure on NH sensors based on exceptional points due to the added periodicity of the underlying Jordan block structure.

\begin{acknowledgments}

This work has been supported by the National Natural Science Foundation of China (grants No.~W2541020, No.~12274059, No.~12574528, and No.~1251101297).

\end{acknowledgments}

\appendix

\section{Spectral topological phase transition}
\label{app:spectral}

Here we present the derivation of the condition for the spectral topological phase transition written as Eq.~\eqref{eq:pgc} in the main text. 
We recall that the general 1D tight-binding NH Hamiltonian with OBC is written as  
\begin{align}
    H = & \sum_{n=1}^{L-1} \, \sum_{\mu=1}^{r-1} \, \sum_{\nu=1}^{d-1} \Big[J_0 \ket{n, \mu, \nu} \bra{n, \mu, \nu{+}1} + \text{H.c.} \Big] \nn
    + & \Big[J_L \ket{n, \mu, d} \bra{n, \mu{+}1, 1} + J_R \ket{n, \mu{+}1, 1} \bra{n, \mu, d} \Big] \nn
    + & \Big[J_m \ket{n, r, d} \bra{n{+}1, 1, 1} + J'_m \ket{n{+}1, 1, 1} \bra{n, r, d} \Big],
\end{align}
For PBC, the module index $n$ run from 1 to $L$, where $n {=} L+1$ is equivalent to $n {=} 1$.
The generalized Brillouin zone (GBZ) provides a theoretical formalism to understand the edge localization of the eigenstates due to skin effect in NH systems in the presence of a point gap~\cite{Yao2018Edge, Yokomizo2019Non}.
Now, following the GBZ formalism, the ansatz for an eigenstate of $H$ is written as
\begin{align}
    \ket{\psi} &= \sum_{\substack{n=1, \dots, L \\ \mu=1, \dots, r \\ \nu=1, \dots, d}} \beta^n c_{\mu, \nu} \ket{n, \mu, \nu},
\end{align}
with $\beta$ denoting the localization parameter.
Now the eigen-equation $H \ket{\psi} = E \ket{\psi}$ for a module in the bulk leads to
\begin{align}
    \begin{pmatrix} 
    -E & J_0 &  &  &  &  &  &  &  & \frac{J'_m}{\beta}  \\ 
    J_0 & -E & J_0 \\
    &  &  & \ddots \\
    &  &  & J_0 & -E & J_L \\
    &  &  &  & J_R & -E & J_0 \\
    &  &  &  &  & J_0 & -E & J_0 \\  
    &  &  &  &  &  &  &  & \ddots \\
    J_m \beta  &  &  &  &  &  &  &  & J_0 & -E \\
    \end{pmatrix}    
    \begin{pmatrix} 
    c_{1,1} \\
    c_{1,2} \\
    \vdots \\
    c_{1,d} \\
    c_{2,1} \\
    c_{2,2} \\
    \vdots \\
    c_{r,d}
    \end{pmatrix}    
    = 0 .
    \label{eq:psi}
\end{align}
The condition for non-trivial solutions demands the determinant $\Delta$ of the $rd \times rd$ matrix on the left to be zero. 
To find this determinant, let us recall the Leibnitz formula for the determinant of a square matrix $\textbf{A}$ as $\textrm{Det}[\textbf{A}]{=}\sum_{\sigma \in S_n} \textrm{sgn}(\sigma) \prod_{i}\textbf{A}_{\sigma(i), i}$ where $\textrm{sgn}(\sigma)$ is the sign function of permutation $\sigma$ applied to the i-th row. 
Applying this formula to the matrix above yields the expression of the determinant $\Delta$ to be given by the following expression 
\begin{align}
    \Delta = \Delta' - J_m J_{m}' \Delta'' + (-1)^{rd-1} \left[ \beta J_m J_0^{r(d-1)}J_{L}^{r-1} + \frac{J_m'}{\beta} J_0^{r(d-1)}J_{R}^{r-1}  \right] .
    \label{eq:quadratic}
\end{align}
Here $\Delta'$ is the determinant of the full matrix without corner elements which is explicitly given by 
\begin{align}
    \Delta' = \frac{(F_d - \lambda_{-})\lambda_{+}^{r} + (\lambda_{+} - F_d)\lambda_{-}^{r}}{\sqrt{\tau^2 - 4 J_LJ_RJ_0^{2(d-1)}}}
\end{align}
where $F_d = J_0^{d} U_d\left(\frac{-E}{2J_0}\right), ~ \tau = J_0^{d} U_d\left(\frac{-E}{2J_0}\right) - J_LJ_R J_0^{d-2} U_{d-2}\left(\frac{-E}{2J_0}\right)$, and 
\begin{widetext} 
\begin{align}   
    \lambda_{\pm} =  \frac{J_{0}^{d} U_{d}\left(\frac{-E}{2J_0}\right) - J_L J_R J_{0}^{d-2} U_{d-2}\left(\frac{-E}{2J_0}\right) \pm \sqrt{\left[J_0^d U_{d}\left(\frac{-E}{2J_0}\right) -J_L J_R J_0^{d-2} U_{d-2}\left(\frac{-E}{2J_0}\right)\right]^2 - 4 J_L J_R J_0^{2(d-1)} }}{2}\\
\end{align}
\end{widetext}
Furthermore, $\Delta''$ is the determinant of the largest inner submatrix without corner elements which is explicitly given by  
\begin{align}
    \Delta'' = \frac{(c_1-\lambda_-c_0)\lambda_+^{\,r-2} + (\lambda_+c_0-c_1)\lambda_-^{\,r-2}}{\lambda_+-\lambda_- }
\end{align}
where $c_0 = -E F_{d-2}F_{d-1} -J_L J_R F_{d-2}^2 -J_0^2 F_{d-3}F_{d-1}$ and $c_1 = E^2F_{d-2}F_{d-1}^2 + E J_L J_R F_{d-2}^2F_{d-1} + EJ_0^{2} F_{d-2}^2F_{d-1} + J_L J_R J_0^2 F_{d-2}^3 + E J_0^2 F_{d-3}F_{d-1}^2 + J_L J_R J_0^2 F_{d-3}F_{d-2}F_{d-1}$.
Here, $U_d(.)$ are Chebyshev polynomials of the second kind. To obtain the formulas of $\Delta', \Delta''$ quoted above, we note these determinants can be recursively written in terms of determinants of smaller submatrices of the same circulant tridiagonal form, and subsequently the recursions can be solved via a transfer matrix method. The modular property operationally translates into the structure of the monodromy matrix that one constructs from individual transfer matrices. See Ref.~\cite{teschl2000jacobi} for more mathematical details and similar problems. 
Note that $\Delta', \Delta''$ are independent of $\beta$, hence the condition for vanishing determinant $\Delta$ can be written down as a quadratic in $\beta$ can be written down as,
\begin{align}
    a \beta^2 + b \beta + c = 0 ,
    \label{eq:beta}
\end{align}
with
\begin{align}
    a = J_0^{r(d-1)} \, J_L^{r-1} \, J_m, \qquad c = J_0^{r(d-1)} \, J_R^{r-1} \, J'_m .
\end{align}
Therefore the solutions $\beta_{1}$ and $\beta_{2}$ (for a given value of $E$) satisfy 
\begin{align}
    \beta_{1} \beta_{2} = \frac{J_R^{r-1} \, J'_m}{J_L^{r-1} \, J_m}.
    \label{eq:beta12}
\end{align}
Now following the results derived in Ref.~\cite{Yokomizo2019Non}, we can write $|\beta_{1}| = |\beta_{2}|$.
This emerges as the condition for the bulk-edge correspondence which requires the correct construction of the GBZ based on the trajectories of $\beta_{1}$ and $\beta_{2}$ with the constraint $|\beta_{1}| = |\beta_{2}|$.
This leads to equivalence of the continuum band spectrum with PBC with the spectrum of the chain with OBC in the thermodynamic limit.
In the following, we provide another proof of this condition by expanding on the arguments in Ref.~\cite{Yao2018Edge}.

The solutions $\beta_{1}$ and $\beta_{2}$ can be used to write down the eigenstate as a superposition
\begin{align}
    \ket{\psi} = \sum_{\substack{j=1, 2 \\n=1, \dots, L \\ \mu=1, \dots, r \\ \nu=1, \dots, d}} \alpha_j \, \beta_j^n \, c^{(j)}_{\mu, \nu} \ket{n, \mu, \nu}  \equiv \sum_{\substack{j=1, 2 \\n=1, \dots, L \\ \mu=1, \dots, r \\ \nu=1, \dots, d}} \beta_j^n \, \phi_{j, \mu, \nu} \ket{n, \mu, \nu},
    \label{eq:state}
\end{align}
where we first write the coefficients as $c^{(j)}_{\mu, \nu}$ to stress the dependence on $\beta_j$, and then define $\phi_{j, \mu, \nu} = \alpha_j \, c^{(j)}_{\mu, \nu}$.
Using Eq.~\eqref{eq:psi}, we can write the linear coupled equations for $\phi_{j, \mu, \nu}$ in the bulk of the chain as
\begin{align}
    J_0 \phi_{j, 1, 2} + \frac{J'_m}{\beta_j} \phi_{j, r, d} &= E \phi_{j, 1, 1} \nn
    J_0 \phi_{j, 1, 1} + J_0 \phi_{j, 1, 3} &= E \phi_{j, 1, 2} \nn
    & \vdots \nn
    J_0 \phi_{j, 1, d-1} + J_L \phi_{j, 2, 1} &= E \phi_{j, 1, d} \nn
    J_R \phi_{j, 1, d} + J_0 \phi_{j, 2, 2} &= E \phi_{j, 2, 1} \nn
    & \vdots \nn
    J_0 \phi_{j, r, d-2} + J_0 \phi_{j, r, d} &= E \phi_{j, r, d-1} \nn
    J_m \beta_j \phi_{j, 1, 1} + J_0 \phi_{j, r, d-1} &= E \phi_{j, r, d} .
    \label{eq:phi}
\end{align}
Now we use Eq.~\eqref{eq:state} for the eigen-equation at the left and right edges of the left-most and right-most modules, respectively, and use Eq.~\eqref{eq:phi} to simplify the results.
Specifically, for $(n, \mu, \nu) = (1, 1, 1)$, Eq.~\eqref{eq:state} leads to
\begin{align}
    & J_0(\beta_{1} \phi_{1, 1, 2} + \beta_{2} \phi_{2, 1, 2}) = E (\beta_{1} \phi_{1, 1, 1} + \beta_{2} \phi_{2, 1, 1}) \nn
    \implies & \phi_{1, r, d} + \phi_{2, r, d} = 0 \qquad (\text{using the first equation of Eq.~\eqref{eq:phi}}) .
    \label{eq:left}
\end{align}
Similarly, for $(n, \mu, \nu) = (L, r, d)$, we get
\begin{align}
    & J_0(\beta_{1}^{L} \phi_{1, r, d-1} + \beta_{2}^{L} \phi_{2, r, d-1}) = E (\beta_{1}^{L} \phi_{1, r, d} + \beta_{2}^{L} \phi_{2, r, d}) \nn
    \implies & \frac{\beta_{1}^{L+1}}{\beta_{2}^{L+1}}= -\frac{\phi_{2, 1, 1}}{\phi_{1, 1, 1}} \qquad (\text{using the last equation of Eq.~\eqref{eq:phi}}) .
    \label{eq:right}
\end{align}
Now, using the last equation of Eq.~\eqref{eq:phi}, we can write $\phi_{j, r, d-1}$ in terms of $\phi_{j, 1, 1}$ and $\phi_{j, r, d}$.
This in turn lets us write $\phi_{j, r, d-2}$ in terms of $\phi_{j, 1, 1}$ and $\phi_{j, r, d}$ by using the second last equation of Eq.~\eqref{eq:phi}.
Recursively moving upwards, we can write $\phi_{j, 1, 2}$ in terms of $\phi_{j, 1, 1}$ and $\phi_{j, r, d}$.
Now, using the first equation of Eq.~\eqref{eq:phi}, we see that $\phi_{j, 1, 1}$ is proportional to $\phi_{j, r, d}$.
Taking the proportionality factor as $f_j(J_0, J_L, J_R, J_m, J'_m, E, \beta_j)$, we rewrite Eq.~\eqref{eq:right} as 
\begin{align}
    \frac{\beta_{1}^{L+1}}{\beta_{2}^{L+1}}= -\frac{f_2 \phi_{2, r, d}}{f_1 \phi_{1, r, d}} = \frac{f_2}{f_1}\qquad (\text{using Eq.~\eqref{eq:left}}) ,
\end{align}
where the right hand side is non-zero in general.
Now if $|\beta_{1}| \neq |\beta_{2}|$, then for $|\beta_{1}| < |\beta_{2}|$, the left hand side goes to zero in the thermodynamic limit.
This contradiction leads to the conclusion that indeed, $|\beta_{1}| = |\beta_{2}|$.

Therefore, using Eq.~\eqref{eq:beta12}, we see that 
\begin{align}
    |\beta_{j}| = \sqrt{\left| \frac{J_R^{r-1} \, J'_m}{J_L^{r-1} \, J_m} \right|} .
    \label{eq:beta_amp}
\end{align}
This implies that the point-gap closure occurs for $\left| \frac{J_R^{r-1} \, J'_m}{J_L^{r-1} \, J_m} \right| = 1$.

\section{Band topological phase transition}
\label{app:band}

Here we present the derivation of the band-gap closing points reported in Sec.~\ref{sec:line} for the Bloch Hamiltonian $H_k$ and the generalized Bloch Hamiltonian $H_{\beta}$.
We recall that for $d{=}2$ and $r{=}2$, the Bloch Hamiltonian is given by
\begin{align}
    H_k = 
    \begin{pmatrix} 
    0 & J_0 &  0  & J'_m e^{-ik} \\ 
    J_0 & 0 & J_L & 0 \\
    0 & J_R & 0 & J_0 \\
    J_m e^{ik} & 0 & J_0 & 0 \\
    \end{pmatrix} .
\end{align}
The eigen-spectrum is given by
\begin{align}
    E_k = \pm \sqrt{\frac{\eta_1 \pm \sqrt{\eta_1^2 - 4 \eta_2}}{2}}
    \label{eq:Hk-eig}
\end{align}
where $\eta_1 {=} 2J_0^2 {+} J_m J'_m {+} J_L J_R; \eta_2 {=} (J_0^2 {-} J_L J_m e^{ik})(J_0^2 {-} J'_m J_R e^{-ik})$. The gap closing conditions corresponding to middle two bands, i.e., at zero-energy eigenvalues are subsequently given by (i) $J_0^2 {=} J_L J_m e^{ik}$ , (ii) $J_0^2 {=} J_R J'_m e^{-ik}$.  
The gap closing condition for the other two bands are given by $\eta_1 {=} {\pm }2\sqrt{\eta_2}$.

For the gap closing of the central bands in the example presented in Sec.~\ref{sec:line}, the first condition does not apply and the second condition gives two solutions for $k{=}0$ at $J_R/J_L {\approx} 0.6008, -2.6008$.
The gap closing between the other two bands has a solution for $k{=}\pi$ at $J_R/J_L {\approx} -1.3822$.

Similarly the generalized Bloch Hamiltonian is given by
\begin{align}
    H_{\beta} = 
    \begin{pmatrix} 
    0 & J_0 &  0  & \frac{J'_m}{\beta} \\ 
    J_0 & 0 & J_L & 0 \\
    0 & J_R & 0 & J_0 \\
    J_m \beta & 0 & J_0 & 0 \\
    \end{pmatrix} .
\end{align}
Here the eigenspectrum and gap closing conditions are given by replacing $e^{\pm ik}$ with the generalized Bloch momentum parameter $\beta^{\pm1}$ in the expressions for eigenspectrum and gap closings above. 
Considering $\beta {=} \sqrt{\frac{J_R \, J'_m}{J_L \, J_m}} e^{i \phi}$, we can write the gap closing condition between the central two bands as $J_0^2 {=} \sqrt{J_L J_m J_R J'_m } e^{\pm i \phi}$.
For $\phi{=}0$, we obtain the gap closing points at $J_R/J_L {\approx} 0.3468, -2.3468$, which are close to the original line gap closing points.
For $\phi{=}\pi$, we obtain the additional gap closing points at $J_R/J_L {\approx} -0.5685, -1.4315$.
These are the values at which the winding number jumps between 0 and 1 in Fig.~\ref{fig:line}(i).
The gap closing condition between the central and side bands has a solution for $\phi{=}\pi/2$ at $J_R/J_L {=} -1.75$.

We now determine the line gap closing condition in absence of point gap closing, as exemplified in Sec.~\ref{sec:chiral} with the chiral and inversion symmetric NH SSH model.
From the analysis of the Bloch Hamiltonian, the gap closing between the two central bands occurs at $\frac{J_R}{|J_0|} {=} \frac{1}{J/|J_0|}$ for $k {=} \pm 2\phi$ where $\phi$ is the phase of $J_0$.
On the other hand, the gap closing condition between the central and side bands is given by $J_R {=} \pm J$ for $k {=} \pi$.

\section{Numerical details}
\label{app:numerics}

\begin{figure}[t]
\centering
\includegraphics[width=\linewidth]{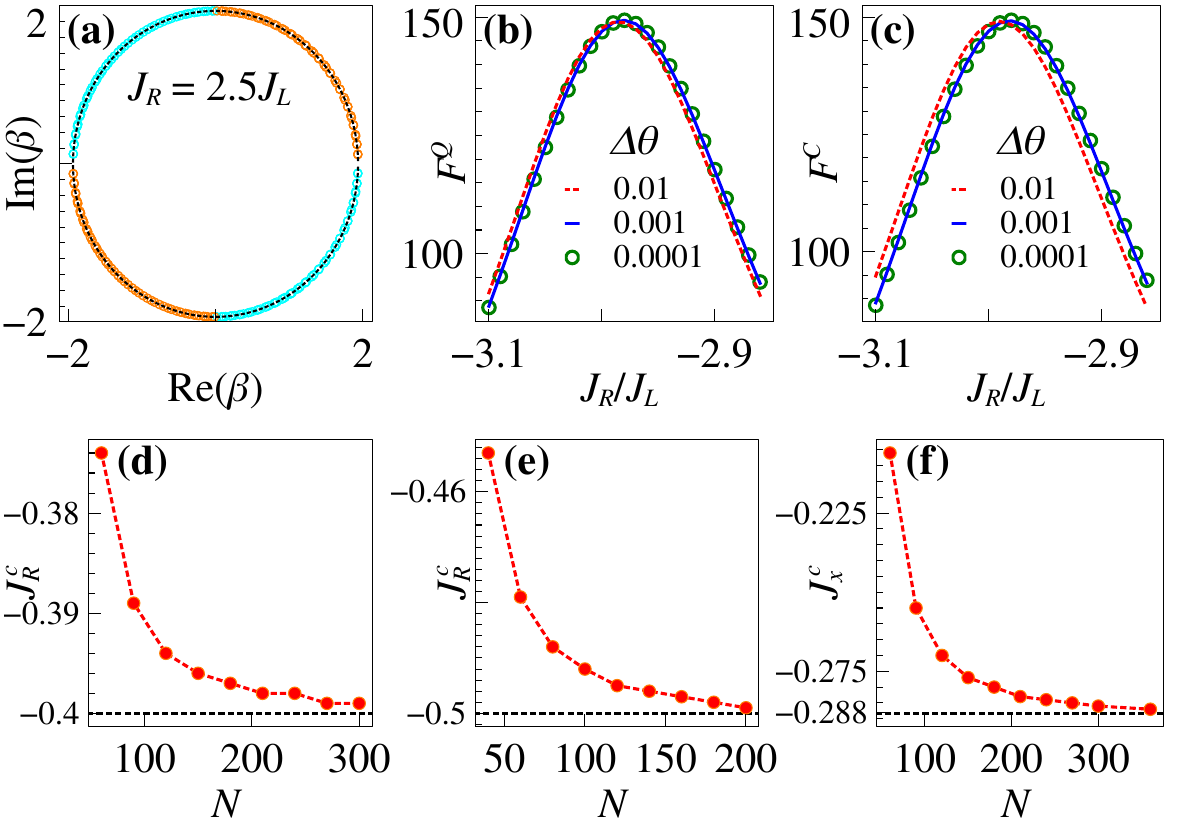} 
\caption{\textbf{Numerical analysis details}. 
(Top panel) Here we exemplify some of the numerical details for $d{=}2, r{=}2, L{=}50, J_0{=}1.25J_L$ with $J_m {=} J_L{+}J, J'_m {=} J_R{+}J$ and $J{=}2J_L$.
(a) Construction of the GBZ as the trajectory of $\beta$ for $J_R{=}2.5J_L$.
The analytical $\beta$ values are shown with black dashed circle with radius $\sqrt{15/4}$, on which the two numerically calculated solutions $\beta_1$ and $\beta_2$ are shown to be residing as the cyan and orange circles. 
Considering the parameter to be estimated $\theta {=} J_R/J_L$, the effect of the step size $\Delta\theta$ on the convergence of the (b) QFI and (c) CFI.
(Bottom panel) The spectral topological phase transition points in finite size systems are shown to approach the analytical value (black dashed lines) as system size $N$ is increased.
(d) Here we consider $d{=}1, r{=}3, L{=}50, J_m {=} J {=} 1/J'_m$ with $J{=}0.4J_L$, for which the asymptotic value of the transition point $J_R^c$ is $-0.4J_L$. 
(e) Here we consider $d{=}2, r{=}2, L{=}50, J_0{=}2J_L, J_m {=} J_L{+}J, J'_m {=} J_R{+}J$ with $J{=}-0.5J_L$, for which the asymptotic value of the transition point $J_R^c$ is $-0.5J_L$.
(f) Here we consider $d{=}1, r{=}3, J_m {=} J {=} 1/J'_m$, with $J{=}J_L/\sqrt{3}$.
For $J_R {=} J_x {+} i J_y$ with $J_y{=}J_L/2$, the asymptotic value of transition point $J_x^c$ is $-J_L/\sqrt{12}$.
}
\label{fig:numerics}
\end{figure}

Constructing the GBZ purely numerically involves obtaining the coefficients of the quadratic equation written down as Eq.~\eqref{eq:beta}.
These coefficients depend on the system parameters as well as the eigen-energies which are first collected through exact diagonalization of the Hamiltonian.
Each eigen-energy therefore gives two $\beta$ values.
It is possible to obtain some pairs of solutions $\beta_1, \beta_2$ which do not satisfy the criterion $|\beta_{1}| = |\beta_{2}|$, and need to be discarded.
The trajectory of all the legitimate $\beta$ values thus obtained gives the GBZ.
The validity of the numerical approach can be checked with the analytical prediction $|\beta| = \sqrt{\left| \frac{J_R^{r-1} \, J'_m}{J_L^{r-1} \, J_m} \right|}$.
We showcase this in Fig.~\ref{fig:numerics}(a) by plotting the numerically calculated $\beta$ values on top of the analytical solution.

Computing QFI and CFI requires the calculation of the steady state and obtaining its partial derivative with respect to the parameter to be estimated.
We use exact diagonalization to determine the steady state and ensure that the norm is one.
The steady state corresponds to the eigenstate with the largest imaginary part in the eigen-energy.
In some cases, there are two eigenstates with largest imaginary energy, with equal and opposite real energies.
We choose the negative real energy for the steady state as our numerical check shows that the QFI near the transition behaves similarly for both the states.
The derivatives with respect to parameter $\theta$ appearing in the expressions of QFI and CFI are calculated with forward difference method to approximate $\ket{\partial_{\theta} \psi} {\approx} \frac{\ket{\psi(\theta+\Delta \theta)} - \ket{\psi(\theta)}}{\Delta \theta}$.
As shown in Fig.~\ref{fig:numerics}(b) and (c), convergence is obtained for $\Delta \theta$ typically in the range $10^{-4} - 10^{-3}$.
We usually consider $\Delta \theta {=} 10^{-5}$ in our calculations, as it is not numerically costly.

To address the shift in the spectral topological phase transition point with change in the size of finite systems, we perform checks on the different cases considered in this work. 
The bottom panel of Fig.~\ref{fig:numerics} shows how the asymptotic value of the transition point is approached as the system size increases for the specific models showcased in Sec.~\ref{sec:sensor} for sensing purposes.

\begin{table}[t]   
\centering
\begin{tabular}{| c | c | c | c | c | c |}

\hline
Case & $N_{\rm max}$ & \multicolumn{2}{|c|}{Modular} & \multicolumn{2}{|c|}{Non-modular} \\ 
\cline{3-6} & & $b$ & $\sigma_b$ & $b$ & $\sigma_b$ \\

\hline
\multirow{3}{*}{Fig.~4c} & $240$ & $2.001$ & $2.421\times10^{-5}$  & $1.988$ & $3.815\times10^{-4}$ \\
\cline{2-6} 
& $1200$ & $2.001$ & $1.814\times10^{-5}$  & $1.994$ & $3.896\times10^{-4}$ \\
\cline{2-6}
& $2400$ & $2.001$ & $1.755\times10^{-5}$  & $1.995$ & $2.774\times10^{-4}$ \\

\hline
\multirow{3}{*}{Fig.~4g} & $240$ & $1.989$ & $3.426\times10^{-4}$  & $1.990$ & $3.184\times10^{-4}$ \\
\cline{2-6} 
& $1200$ & $1.995$ & $3.494\times10^{-4}$  & $1.994$ & $3.205\times10^{-4}$ \\
\cline{2-6}
& $2400$ & $1.996$ & $2.494\times10^{-4}$  & $1.996$ & $2.273\times10^{-4}$ \\

\hline

\end{tabular}
\caption{\textbf{Fitting details}. The scaling exponents $b$ and the associated uncertainties in them $\sigma_b$ for the various cases considered in the main text, with different maximum system size $N_{\rm max}$ considered while keeping the lowest size fixed at 120.}
\label{tab:exponent}
\end{table}

The fitting exponents, required for the scaling analysis of Fisher information, are obtained using the curve\_fit function of the optimize package of scipy library in Python.
One can get the uncertainties in the exponents by taking the square root of the corresponding diagonal elements of the generated covariance matrix output.
As shown in Table~\ref{tab:exponent} for the cases considered in main text for Figs.~\ref{fig:single}(c) and (g), the exponents usually approach the quadratic value as the maximum system size $N_{\rm max}$ is increased from a few hundreds to a few thousands, while the numerical uncertainties gradually decreasing within a range of $10^{-5} - 10^{-4}$.

\bibliography{Ref}

\end{document}